\documentclass[12pt]{iopart}

\usepackage{framed}
\usepackage{graphicx}
\usepackage{epsfig}

\newcommand{\ud}{\/\mathrm{d}\/}
\newcommand{\rhom}{\rho_{\mathrm{m}}}
\newcommand{\JH}{J_\mathrm{H}}
\newcommand{\Ptwo}{P^{(2)}}
\newcommand{\nub}{\nu_{\mathrm{B}}}

\begin{document}

\title[Stochastic energy exchange models of locally
  confined hard spheres]
{Heat transport in stochastic energy exchange models of locally
  confined hard spheres}
\author{Pierre Gaspard and Thomas Gilbert}
\address{Center for Nonlinear Phenomena and Complex Systems,\\
  Universit\'e Libre  de Bruxelles, C.~P.~231, Campus Plaine, B-1050
  Brussels, Belgium}

\eads{\mailto{gaspard@ulb.ac.be}, \mailto{thomas.gilbert@ulb.ac.be}}

\begin{abstract}
We study heat transport in a class of stochastic energy exchange systems
that characterize the interactions of networks of locally trapped hard
spheres under the assumption that neighbouring particles undergo rare
binary collisions. Our results provide an extension to three-dimensional
dynamics of previous ones applying to the dynamics of confined
two-dimensional hard disks [Gaspard P \& Gilbert T \emph{On the derivation
  of   Fourier's law in stochastic energy exchange systems} J Stat Mech
(2008) P11021]. It is remarkable that the heat conductivity is here again
given by the frequency of energy exchanges. Moreover the expression of the
stochastic kernel which specifies the energy exchange dynamics is simpler
in this case and therefore allows for faster and more extensive numerical
computations.
\end{abstract}

\submitto{J. Stat. Mech.}

\section{Introduction \label{sec.intro}}

The derivation of Fourier's law of heat transfer in insulating solid
materials is a difficult problem which has been challenging theoretical
physicists for close to two centuries \cite{BLRB00}. However, recent works
\cite{GG08a, GL08} on models of confined particles in interaction have shed
new light on this problem, setting the stage for a systematic derivation of
Fourier's law from first principles. These works suggest that there
is hope to achieve such a first principles characterization of heat
transfer and prove the validity of Fourier's law in a particular class of
insulating materials known as aerogels.

In order to achieve such a characterization, the authors of \cite{GG08a,
  GL08} have focused their attention on classes of models which, like
aerogels, combine the collisional dynamics of gases with the spatial
structure of solids. Starting from a Hamiltonian description, it was  
shown that the heat conductivity of such models is universally given by the
frequency of collisions between gas particles. This universality manifests
itself when the gas particles are individually trapped in a porous solid
material and only rarely collide with each other, mostly rattling around
their traps. Under this assumption, individual particles typically achieve
local equilibrium states at their respective kinetic energies, ergodically
exploring their trapping cells, before energy exchanges proceed. 

This local equilibration mechanism is rather similar to the trapping
mechanism of tracer particles in a periodic Lorentz gas which lead Machta
and Zwanzig \cite{MZ83} to infer a stochastic approximation of mass
transport in the Lorentz gas. In this approximation, the diffusion
coefficient is identified, up to dimensional factors, with the rate of jump
of tracer particles from cell to cell.  

Likewise, in our systems, this local equilibration naturally yields a
stochastic description of the time evolution of the probability
distribution of the local energies in terms of a master equation. Such
derivation was extensively studied in \cite{GG08b, GG08c}, relating to the
dynamics of confined two-dimensional hard-disks. Irrespective of the
dimensionality of the underlying dynamics, the main difficulty when
analyzing the transport properties of such stochastic systems is that, 
unlike with the Machta-Zwanzig models which deal with independent tracer
dynamics, the energy distributions cannot be reduced to single cell
distributions. Nevertheless, the transport coefficient --in this case the
heat conductivity-- is identified, up to dimensional factors, as the rate of
energy exchanges. 

The purpose of this paper is to provide an extension of the results
presented in \cite{GG08c} to the dynamics of trapped  three-dimensional
hard spheres. Starting from the Liouville equation which describes the time
evolution of phase-space distributions of such systems, we consider the
reduction of this equation 
to a stochastic evolution for the local energies under the assumption that
collisions between neighbouring particles are rare compared to wall
collisions. We subsequently show that the stochastic kernel which
characterizes the energy exchanges between neighbouring cells has the
symmetries described in \cite{GG08c} which ensure the identity between the
heat conductivity and frequency of energy exchanges. Furthermore, the
expression of the stochastic kernel is much simpler in the case of
three-dimensional hard spheres than in that of two-dimensional hard
disks. The model is thus amenable to higher precision numerical
simulations, which allows us to confirm the preceding arguments to a higher
precision than had been previously obtained in the framework of underlying
two-dimensional dynamics. 

The paper is organized as follows. A summary of the results described in
\cite{GG08c} is presented in section \ref{sec.summary}. In section
\ref{sec.3dsys} we introduce mechanical models of periodic networks of
confined hard spheres and discuss the necessary assumptions upon which
these systems are amenable to a stochastic reduction. The statistical
evolution of these systems is considered in section 
\ref{sec.statevol} and its reduction to a stochastic equation in section
\ref{sec.meq}, with a detailed derivation of the stochastic kernel. The
identity between the energy exchange frequency and the 
thermal conductivity is established in section \ref{sec.kappa}. Numerical
results supporting our theoretical arguments are presented in section
\ref{sec.num}. Conclusions and perspectives are drawn in section
\ref{sec.con}.

\section{Summary of the main results \label{sec.summary}}

Consider a system of $N$ energy cells $\epsilon_1, \dots, \epsilon_N$,
with stochastic exchange of energy among pairs of neighbouring cells. We
assume that the statistical evolution is described by the following master
equation, 
\begin{eqnarray}
  \fl \partial_t P_N(\epsilon_1, \dots, \epsilon_N, t) 
  \nonumber\\
  \lo = \frac{1}{2}\sum_{a,b = 1}^N
  \int \ud\eta \Big[W(\epsilon_a + \eta, \epsilon_b - \eta| \epsilon_a,
  \epsilon_b)
  P_N(\dots, \epsilon_a + \eta, \dots,\epsilon_b - \eta, \dots, t)
  \nonumber\\
  - W(\epsilon_a, \epsilon_b | \epsilon_a - \eta, \epsilon_b + \eta)
  P_N(\dots, \epsilon_a, \dots, \epsilon_b, \dots, t)\Big]\,,
  \label{mastereq}
\end{eqnarray}
where $W(\epsilon_a, \epsilon_b | \epsilon_a - \eta, \epsilon_b + \eta)$ is
the stochastic kernel specifying the process of exchange of energy $\eta$
between two neighbouring cells $a$ and $b$ at respective energies
$\epsilon_a$ and $\epsilon_b$, with $-\epsilon_b \leq \eta \leq
\epsilon_a$. We will be concerned here with kernels that do not depend on
the specific pair $a,b$ of neighbouring cells.

Given the two energies $\epsilon_a$ and $\epsilon_b$, the characteristic
time scale of energy exchanges between neighbouring cells $a$ and $b$ is
determined by the collision frequency
\begin{equation}
  \nu(\epsilon_a, \epsilon_b) \equiv
  \int\ud \eta W(\epsilon_a, \epsilon_b | \epsilon_a - \eta,\epsilon_b +
  \eta)\,,
  \label{collfreq}
\end{equation}
whose canonical equilibrium average at temperature $T$ we denote $\nub(T)
\equiv \langle \nu(\epsilon_a, \epsilon_b) \rangle_T \sim \sqrt{T}$.

In the proper hydrodynamic scaling limit, this master equation yields the
time evolution of the local temperatures, 
\begin{equation}
  T_a(t) = \frac{2}{d}\int \prod_{i = 1}^N \ud \epsilon_i \,
  \epsilon_a P_N(\epsilon_1, \dots, \epsilon_N, t)\,,
\end{equation}
where $d$ is the dimension of the underlying dynamics. This evolution turns
out to be given according to Fourier's law, namely 
\begin{equation}
  \partial_t T(x,t) = - \partial_x[\kappa(T) \partial_x T(x,t)]\,,
  \label{fourierlaw}
\end{equation}
where the heat conductivity $\kappa(T)$ is
\begin{equation}
  \kappa(T) = \nub(T)\,.
  \label{kappaeqnu}
\end{equation}
The result (\ref{kappaeqnu}) establishes an identity between a macroscopic
quantity, the heat conductivity, and a microscopic one, the frequency of
energy exchanges between two neighbouring cells. As shown in \cite{GG08c},
the derivation of this identity follows from equation
(\ref{mastereq}) in two independent ways, which rely on special
symmetries of the kernel. 

These symmetries concern the equilibrium averages of the first and second
moments of the energy exchanges,
\begin{eqnarray}
  j(\epsilon_a, \epsilon_b) \equiv 
  \int\ud \eta \, \eta W(\epsilon_a, \epsilon_b | \epsilon_a -
  \eta,\epsilon_b + \eta)\,,
  \label{current}\\
  h(\epsilon_a, \epsilon_b) \equiv 
  \int\ud \eta \, \eta^2 W(\epsilon_a, \epsilon_b | \epsilon_a -
  \eta,\epsilon_b + \eta)\,.
  \label{etasq}
\end{eqnarray}
In \cite{GG08c}, we showed that the kernel $W$ associated with the
confining dynamics of two-dimensional hard disks has the symmetries:
\begin{equation}
  \left\langle \nu(\epsilon_a, \epsilon_b)\right\rangle_T
  = \frac{1}{2} \left\langle (\epsilon_a - \epsilon_b) 
    j(\epsilon_a, \epsilon_b)\right\rangle_T
  = 
  \frac{1}{2} \left\langle h(\epsilon_a, \epsilon_b)
  \right\rangle_T\,.
  \label{symkernel}
\end{equation}

Given the master equation (\ref{mastereq}), there are two ways of
computing the heat conductivity. The first one is to
assume a non-equilibrium stationary state resulting from a temperature
gradient. One then finds that the heat conductivity has expression
\begin{equation}
  \kappa(T) = \frac{1}{2} \left\langle (\epsilon_a - \epsilon_b) 
  j(\epsilon_a, \epsilon_b)\right\rangle_T\,,
  \label{kappanoeq}
\end{equation}
which, using the properties of the kernel (\ref{symkernel}), in its turn
yields (\ref{kappaeqnu}). 

The second way of computing the heat conductivity is through the Green-Kubo
formula, from which it turns out that only static correlations contribute
to the transport coefficient with expression
\begin{equation}
  \kappa(T) = \frac{1}{2} \left\langle h(\epsilon_a, \epsilon_b)
    \right\rangle_T\,.
  \label{kappaeq}
\end{equation}
And, again using the symmetries of the kernel (\ref{symkernel}), we obtain
(\ref{kappaeqnu}). 

The goal of this paper is to provide a derivation of the master equation
(\ref{mastereq}) associated with the stochastic energy exchanges of
rarely interacting confined hard spheres and show that the corresponding
kernel has the symmetries (\ref{symkernel}). The identity (\ref{kappaeqnu})
ensues.

We show in this paper that these results extend verbatim to the kernel
associated with the confining dynamics of three-dimensional hard spheres.

\section{Networks of confined hard spheres \label{sec.3dsys}}

\begin{figure}[htpb]
  \centering
  \includegraphics[width=.5\textwidth]{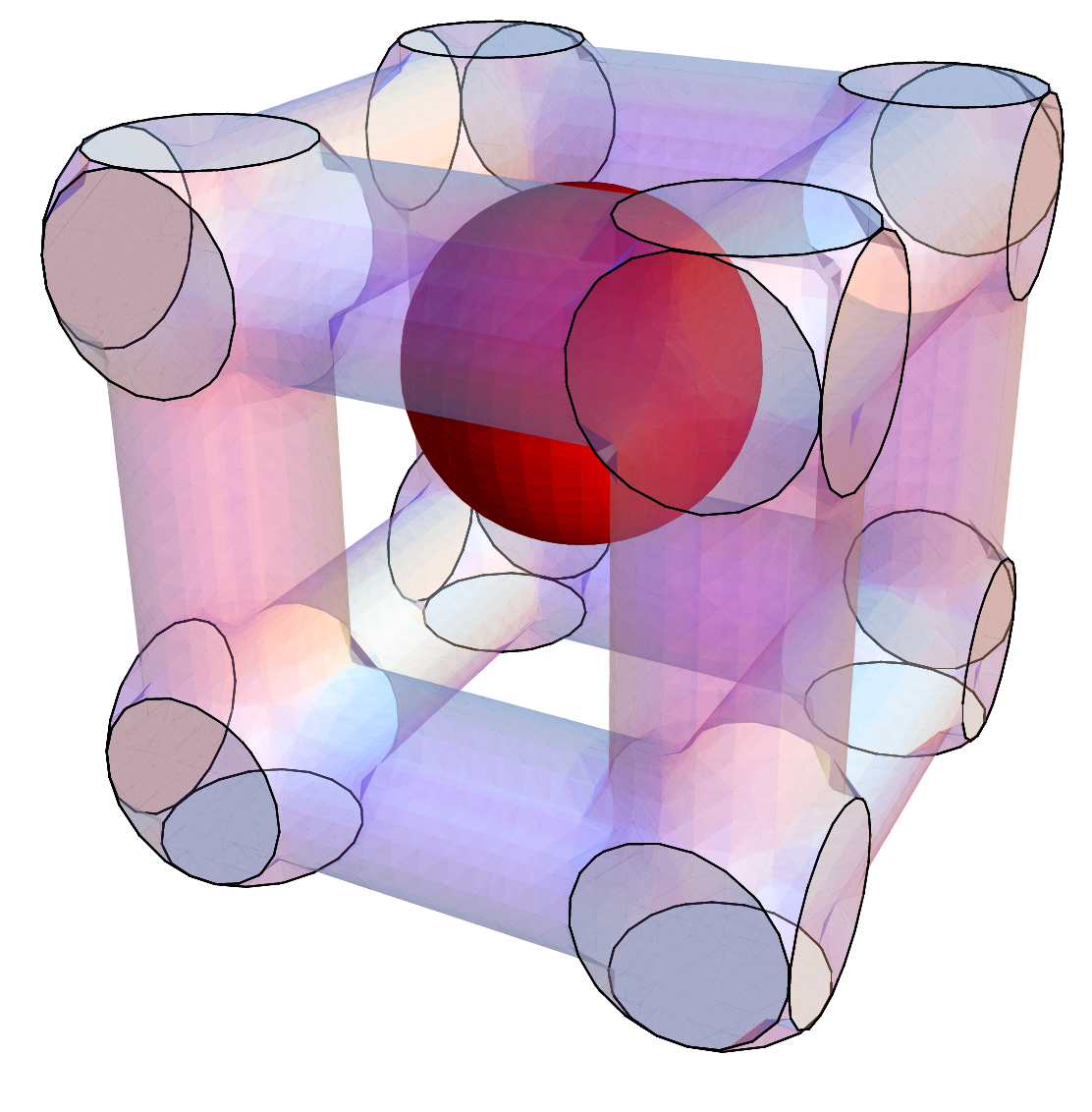}
  \caption{Hard-sphere particle trapped in a cuboid cell with cylindrical
    edges. One imagines a material consisting of many copies of such cells
    forming a spatially periodic structure. The cells are semi-porous in the sense that
    particles are prevented from escaping, and can yet partially
    penetrate into the neighbouring cells, thus allowing energy transfer
    through collisions among neighbouring particles. The likelihood of such
    collision events can be controlled by the geometry of the cell.} 
  \label{fig.example}
\end{figure}

Consider a lattice of confining three-dimensional cells, each containing a
single hard-sphere particle. The mechanism of confinement prevents mass
transport. However we imagine a form of semi-porosity by which particles
in neighbouring cells are able to perform elastic collisions with each
other, thereby exchanging energy. Such a mechanical system with hard-core
confinement is depicted in figure \ref{fig.example}. Two dimensional
versions of similar systems were extensively studied in \cite{GG08a,
  GG08b}. We note that the geometrical details of the confining cells are
irrelevant for what follows so long as the local dynamics mixes the
velocity angles. That is, each cell taken individually with a single moving
particle inside them is a semi-dispersing billiard with uniform equilibrium
measure on the constant kinetic energy surface. This assumption validates
the local equilibrium distributions described in the next section. 

These systems combine the collisional dynamics of gases and the spatial
structure of solids. There is therefore a natural distinction between the
local dynamics, which deal with the interactions between the moving
particles and the solid matrix of their confining cells\footnote{In our
  models, no energy
  exchanges take place between the confining walls and the moving
  particles.}, and the interacting dynamics, by which two moving particles
in neighbouring cells perform an elastic collision. On the one hand, the
local dynamics are  characterized by a wall-collision frequency,
$\nu_\mathrm{W}$, which depends on the geometry of the confining cell as
well as on the kinetic energy of the moving particles through well known
results of ergodic theory \cite{CM06}. On the other hand, the interacting 
dynamics are characterized by the frequency of binary collisions,
\emph{i.e.} the frequency of collisions between neighbouring particles,
denoted $\nu_\mathrm{B}$.

We assume that the scale separation, $\nu_\mathrm{B}\ll\nu_\mathrm{W}$, is
achieved. This involves specific conditions which depend on the geometry of
the confining cells, but which will not concern us here. This assumption
is to say that individual particles typically perform many collisions with
the solid matrix of their confining cells, rattling about their cages at
higher frequency than that of binary collisions. As will be shown in the
next section, in this regime, Liouville's equation governing the time
evolution of phase-space densities reduces to a master equation for the 
time evolution of local energies. The scale separation between the two
collision frequencies $\nu_\mathrm{B}$ and $\nu_\mathrm{W}$ is the only
parameter that controls the validity of this reduction. Moreover it becomes
exact in the limit of vanishing binary collision frequency.

\section{Statistical evolution \label{sec.statevol}}

The phase-space probability density of the mechanical systems of $N$
confined hard spheres is specified by the distribution function
$p_N(\bi{r}_1,\bi{v}_1,\dots,\bi{r}_N,\bi{v}_N)$, where
$\bi{r}_a$ and $\bi{v}_a$, $a = 1,\dots, N$, denote the $a$th
particle position and velocity vectors. The index $a$ stands for the label
of the confining cells. For our system, as is customary for hard sphere
dynamics,  this distribution satisfies a pseudo-Liouville equation
\cite{EDHVL69}, which is well defined in spite of the singularity of the
hard-core interactions. This equation, which describes the time evolution
of $p_N$ is composed of three types of terms: (i) the advection terms,
which account for the displacement of the moving particles within their
respective billiard cells; (ii) the wall collision terms, which account for
the wall collision events, between the moving particles and the solid
matrix of their confining cells; and (iii) the binary collision terms,
which account for binary collision events, between moving particles
belonging to neighbouring cells~:
\begin{equation}
\partial_t p_N = \sum_{a = 1}^N \left[
-\bi{v}_a \cdot \partial_{\bi{r}_a}
+ K^{(a)}\right] p_N
+ \frac{1}{2}\sum_{a,b=1}^N B^{(a,b)} p_N\,.
\label{pseudoliouville}
\end{equation}
The first two terms account for the motion of individual particles
within their respective cells, whether free advection or wall collisions,
and the third term is the binary collision operator which, 
written in terms of the relative  positions $\bi{r}_{ab}$ and velocities
$\bi{v}_{ab}$ of particles $a$ and $b$, and the unit vector
$\hat{\bi{e}}_{ab}$ that connects them, is 
\begin{eqnarray}
\lo
B^{(a,b)} p_N(\dots,\bi{r}_a, \bi{v}_a,\dots,
\bi{r}_b, \bi{v}_b, \dots)
=\nonumber\\
(2 \rhom)^2 \int_{\hat{\bi{e}}_{ab}\cdot\bi{v}_{ab}>0}
\ud\hat{\bi{e}}_{ab}(\hat{\bi{e}}_{ab}\cdot\bi{v}_{ab})
\Big[\delta(\bi{r}_{ab} - 2\rhom \hat{\bi{e}}_{ab})
\nonumber\\
\times p_N\big(\dots,\bi{r}_a, \bi{v}_a -
\hat{\bi{e}}_{ab}(\hat{\bi{e}}_{ab}\cdot\bi{v}_{ab}), \dots,
\bi{r}_b,  \bi{v}_b +
\hat{\bi{e}}_{ab}(\hat{\bi{e}}_{ab}\cdot\bi{v}_{ab})
,\dots\big) \nonumber\\
- \delta(\bi{r}_{ab} + 2\rhom\hat{\bi{e}}_{ab})
p_N\big(\dots, \bi{r}_a, \bi{v}_a,\dots,\bi{r}_b, \bi{v}_b,
\dots\big) \Big]\,.
\label{binarycollisions}
\end{eqnarray}

\subsection{Equilibrium states}

Equilibrium states of equation (\ref{pseudoliouville}) are product
measures such as a Maxwellian distribution with inverse temperature
$\beta$, 
\begin{equation}
  p_N^{(\mathrm{can})}(\bi{r}_1,\bi{v}_1,\dots,\bi{r}_N,\bi{v}_N) 
  =   \frac{1}{|\mathcal{L}(N)|}
  \left(\frac{m\beta}{2\pi}\right)^{\frac{3N}{2}}
  e^{-\frac{m}{2}\beta(v_1^2+ \dots + v_N^2)}\,,
  \label{caneq}
\end{equation}
where $|\mathcal{L}(N)|$ denotes the volume accessible to all $N$
particles, or micro-canonical measures of totally energy $E$,
\begin{equation}
  p_N^{(\mathrm{mic})}(\bi{r}_1,\bi{v}_1,\dots,\bi{r}_N,\bi{v}_N) 
  =   \frac{\Gamma(\frac{3N}{2})E}{|\mathcal{L}(N)|} 
  \left(\frac{m}{2\pi E}\right)^{\frac{3N}{2}}
  \delta\left(\frac{m}{2}\sum_a v_a^2 - E\right)\,,
  \label{miceq}
\end{equation}
where the Gamma function is
\begin{equation}
  \Gamma\left(\frac{3N}{2}\right) = 
  \left\{
    \begin{array}{l@{\quad}l}
      (3N-2)!!\sqrt{2\pi}\,2^{-\frac{3N-2}{2}}\,,&N\,\mathrm{odd}\,,\\
      (3N-2)!!2^{-\frac{3N-2}{2}}\,,&N\,\mathrm{even}\,.\\
      \end{array}
    \right.
\end{equation}

\subsection{Local equilibrium states}

Under the assumption that particles typically collide more often with the
solid matrix of their confining cells than among nearest neighbours, the
time-evolution (\ref{pseudoliouville}) is dominated by the local terms so
that the distribution $p_N$ rapidly evolves toward a local equilibrium
distribution, which depends upon the energy variables
$\{\epsilon_1,\dots,\epsilon_N\}$ only.

We denote by $P_N(\epsilon_1,\dots,\epsilon_N, t)$ this object, invariant
under the local evolution terms, defined in terms of the phase-space
distribution function of the mechanical systems of $N$ confined particles
$p_N(\bi{r}_1,\bi{v}_1,\dots,\bi{r}_N,\bi{v}_N, t)$ according to
\begin{equation}
  \fl
  P_N(\epsilon_1, \dots, \epsilon_N, t) \equiv
  \int \prod_{a=1}^N\ud \bi{r}_a\ud \bi{v}_a 
  \delta \left(\epsilon_a - \frac{m}{2} v_a^2 \right)
  p_N(\bi{r}_1,\bi{v}_1,\dots,\bi{r}_N,\bi{v}_N,t)\,.
\label{ple}
\end{equation}
Notice that the normalization of $p_N$ implies that of $P_N$.

In terms of the energy distributions, the corresponding equilibrium
measures are
\begin{eqnarray}
  \fl
  P_N^{(\mathrm{can})}(\epsilon_1, \dots, \epsilon_N) =
  \left(\frac{m\beta}{2\pi}\right)^{\frac{3N}{2}}
  \int \prod_{a=1}^N\ud \bi{v}_a 
  \delta \left(\epsilon_a - \frac{m}{2} v_a^2 \right)
  e^{-\frac{m}{2}\beta(v_1^2 + \dots + v_N^2)}\,,
  \nonumber \\
  =   \left(\frac{2\beta^{\frac{3}{2}}}{\sqrt{\pi}}\right)^N
  \sqrt{\epsilon_1\cdots\epsilon_N}
  e^{-\beta(\epsilon_1 + \dots + \epsilon_N)}\,.
  \label{eqplecan}
  \\
  \fl
  P_N^{(\mathrm{mic})}(\epsilon_1, \dots, \epsilon_N) 
  \nonumber\\
  =
  \Gamma\left(\frac{3N}{2}\right) 
  E \left(\frac{m}{2\pi E}\right)^{\frac{3N}{2}}
  \int \prod_{a=1}^N\ud \bi{v}_a 
  \delta\left(\epsilon_a - \frac{m}{2} v_a^2\right)
  \delta\left(\frac{m}{2}\sum_a v_a^2 - E\right)\,,
  \nonumber\\
  =   \Gamma\left(\frac{3N}{2}\right)  
    \left(\frac{2}{\sqrt{\pi}E}\right)^N
  \sqrt{\frac{\epsilon_1\cdots\epsilon_N}{E^N}}
  \delta\left(\frac{\epsilon_1}{E} + \dots + \frac{\epsilon_N}{E} - 1
  \right)\,.
  \label{eqplemic}
\end{eqnarray}

One can check that both these measures are normalized. 
We notice that the one- and two-point distribution of the canonical and 
micro-canonical
measure (\ref{eqplemic}) have the forms ($N\ge 3$)
\begin{eqnarray}
  P_N^{(\mathrm{can})}(\epsilon) = 
  \frac{2\beta^{\frac{3}{2}}}{\sqrt{\pi}}\sqrt{\epsilon} e^{-\beta\epsilon}\,,
  \label{canple1pt}\\
  P_N^{(\mathrm{can})}(\epsilon_a, \epsilon_b) = 
  \frac{4\beta^3}{\pi}\sqrt{\epsilon_a \epsilon_b} 
  e^{-\beta(\epsilon_a + \epsilon_b)}\,,
  \label{canple2pt}\\
  P_N^{(\mathrm{mic})}(\epsilon) = 
    \frac{2}{\sqrt{\pi} E^{\frac{3}{2}}}\frac{\Gamma(\frac{3}{2}N)}
    {\Gamma[\frac{3}{2}(N-1)]} \sqrt{\epsilon}
    \left(1 - \frac{\epsilon}{E}\right)^{\frac{3N-5}{2}}
    \,,\label{micple1pt}\\
  P_N^{(\mathrm{mic})}(\epsilon_a, \epsilon_b) = 
    \frac{4\Gamma(\frac{3}{2}N)}{\pi E^3\Gamma(\frac{3}{2}N-3)}
    \sqrt{\epsilon_a \epsilon_b}
    \left(1 - \frac{\epsilon_a}{E} -
      \frac{\epsilon_b}{E}\right)^{\frac{3N - 8}{2}}
    \,.\label{micple2pt}
\end{eqnarray}
The canonical ensemble marginals are straightforward. As of the
micro-canonical ensemble, the marginals are obtained from the $N$-point
distribution (\ref{eqplemic}) by integrating out $N-2$ of the $N$ energy
variables. The integration over the first of $N-2$ variables takes care of
the delta function. Each subsequent integration has the form
\begin{equation}
  \int_0^x \ud y\,\sqrt{y}(x - y)^a\sim x^{\frac{3}{2} + a}\,,
\end{equation}
where $x = 1 - \epsilon_1/N - \dots - \epsilon_n/N$. The successive values
of $a$ are 
\begin{equation}
  \begin{array}{l@{\quad \mathrm{when}\,\mathrm{integrating}\,\mathrm{over} \quad}l}
    \frac{1}{2}&\,\epsilon_{N-1}\,,\\ 
    2 & \,\epsilon_{N-2}\,,\\ 
    \frac{7}{2} &\,\epsilon_{N-3}\,,\\ 
    5 &\,\epsilon_{N-4}\,,\\ 
    \end{array}
\end{equation}
and so on until we obtain (\ref{micple2pt}) after $N-2$ integrations.

Letting $E = 3/2 N \beta^{-1}$ into (\ref{micple1pt}), and
taking the limit $N\to\infty$, we recover
\begin{equation}
  \lim_{N\to\infty}   P_N^{(\mathrm{mic})}(\epsilon) = 
  \frac{2\beta^{\frac{3}{2}}}{\sqrt{\pi}}\sqrt{\epsilon}\,
  e^{-\beta\epsilon}\,,
\end{equation}
which is the one-particle distribution of the canonical measure
(\ref{eqplecan}). 

The energy moments with respect to the one cell distributions
(\ref{canple1pt}), for the canonical ensemble, and (\ref{micple1pt}), for
the micro-canonical ensemble, are easily computed. Let $\langle
. \rangle_\mathrm{can}$ and $\langle . \rangle_\mathrm{mic}$  
denote the averages with respect to the distributions (\ref{canple1pt}) and
(\ref{micple1pt}) at energy $E = 3/2N\beta^{-1}$ respectively. The
corresponding $n$th moments of the energy are
\begin{eqnarray}
  \langle (\beta\epsilon)^n \rangle_\mathrm{can}
  &= \frac{2}{\sqrt{\pi}} \Gamma\left(\frac{3}{2} + n \right)\,,\nonumber\\
  &= \frac{(2n+1)!!}{2^n}\,,
  \label{canenmoment}\\
  \langle (\beta\epsilon)^n \rangle_\mathrm{mic}
  &= \frac{2}{\sqrt{\pi}}\left(\frac{3N}{2}\right)^n 
  \frac{\Gamma(\frac{3}{2}+n)
    \Gamma(\frac{3}{2}N)}{\Gamma(\frac{3}{2}N + n)}\,,\nonumber\\
  &= \frac{(2n+1)!!}{2^n} \frac{\left(\frac{3N}{2}\right)^n 
    \Gamma(\frac{3}{2}N)}{\Gamma(\frac{3}{2}N + n)}\,.
  \label{micenmoment}
\end{eqnarray}
In particular the first few moments are
\begin{equation}
  \eqalign{
    \langle \beta \epsilon \rangle_\mathrm{mic} = \frac{3}{2}\,,\\
    \langle (\beta\epsilon)^2 \rangle_\mathrm{mic} 
    = \frac{15}{4}\frac{3N}{3N + 2}\,,\\
    \langle (\beta \epsilon)^3 \rangle_\mathrm{mic} 
    = \frac{105}{8}\frac{9 N^2}{(3N + 4)(3N + 2)}\,.\\
  }
  \label{micenfirst}
\end{equation}
Thus only the first one is independent of $N$ and equal to its canonical
counterpart. The remaining ones are only equal to their canonical
counterparts up to $\mathcal{O}(1/N)$ corrections.

\subsection{Local equilibrium closure}

Taking the time-derivative of equation (\ref{ple}) and substituting
equation (\ref{pseudoliouville}), only the term involving energy exchanges
between neighboring particles contribute~:
\begin{eqnarray}
  \fl
  \partial_t P_N(\epsilon_1, \dots, \epsilon_N, t) &= 
  \int \prod_{i=1}^N\ud \bi{r}_i\ud \bi{v}_i\delta \left( \epsilon_i - 
  \frac{m}{2} v_i^2\right)
  \partial_t p_N(\bi{r}_1,\bi{v}_1,\dots,\bi{r}_N,\bi{v}_N,t)
  \,,\nonumber\\
  &=
  \frac{1}{2} \sum_{a,b=1}^N
  \int \prod_{i=1}^N\ud \bi{r}_i\ud \bi{v}_i 
  \delta \left(\epsilon_i - \frac{m}{2} v_i^2 \right)\,
  B^{(a,b)}p_N(\bi{r}_1, \bi{v}_1,\dots, \bi{r}_N, \bi{v}_N, t)
  \label{tevple}
\end{eqnarray}
The action of $B^{(a,b)}$ on $p_N$ is given according to equation
(\ref{binarycollisions}). In order to obtain a closed equation for $P_N$,
we assume that the distribution $p_N$ is a locally 
micro-canonical one, \emph{i.e.} it depends only on the energy distribution
and is thus given according to
\begin{equation}
  \fl
  p_N(\bi{r}_1, \bi{v}_1,\dots, \bi{r}_N, \bi{v}_N, t)
  = \frac{m^N}{(8\pi)^N|\mathcal{L}(N)|}
  \int \prod_{i=1}^N \ud \epsilon_i \epsilon_i^{-1}
  \delta\left( v_i - \sqrt{\frac{2\epsilon_i}{m}}\right)
  P_N(\epsilon_1, \dots, \epsilon_N, t)\,.
  \label{invple}
\end{equation}
The other prefactors are so chosen that this measure is normalized,
\emph{viz.}
\begin{equation}
  \int \prod_{i=1}^N   \ud\bi{r}_i\ud\bi{v}_i 
  p_N(\bi{r}_1, \bi{v}_1,\dots, \bi{r}_N, \bi{v}_N, t)
  = 1\,.
\end{equation}

\section{Master equation \label{sec.meq}}

The validity of equation (\ref{invple}) depends on the scale separation between
the wall and binary collision frequencies which will be assumed throughout
this article. This allows us to focus on the stochastic
reduction of the energy exchange dynamics, considering the mesoscopic level
description of the time evolution of probability densities as a stochastic
evolution.  

Indeed, substituting equation  (\ref{invple}) into
(\ref{binarycollisions}), we write the time evolution of local equilibrium
distributions (\ref{tevple}) in the form of the master equation
(\ref{mastereq}), whose kernel $W$ can be identified from the
computations above. The time evolution is thus specified by a  master
equation which accounts for the energy exchanges between neighbouring
cells and makes no further reference to the collisional dynamics of
confined hard spheres. 

\subsection{Derivation of the master equation}

For each pair $(a,b)$ of neighbouring cells, we have a contribution to
equation (\ref{tevple}) of the form
\begin{eqnarray}
  \fl
  \int \prod_{i=1}^N \ud \bi{r}_i \ud \bi{v}_i 
  \delta\left(\epsilon_i - \frac{m}{2} v_i^2\right) 
  B^{(a,b)} p_N(\{\bi{r}_i, \bi{v}_i\},t) \nonumber\\
  = 4\rhom^2   \int \prod_{i=1}^N \ud \bi{r}_i \ud \bi{v}_i 
  \int_{\hat{\bi{e}}_{ab}\cdot\bi{v}_{ab}>0}
  \ud\hat{\bi{e}}_{ab}(\hat{\bi{e}}_{ab}\cdot\bi{v}_{ab})
  \delta\left(\epsilon_i - \frac{m}{2} v_i^2\right) 
  \label{meqcomp1}
  \\
  \hskip -1cm
  \times\left[ \delta(\bi{r}_{ab} - 2\rhom \hat{\bi{e}}_{ab})
  p_N\big(\dots, \bi{v'}_a, \bi{v'}_b,\dots\big) 
  - \delta(\bi{r}_{ab} + 2\rhom\hat{\bi{e}}_{ab})
  p_N\big(\dots, \bi{v}_a, \bi{v}_b, \dots\big) \right]\,,
\nonumber
\end{eqnarray}
where $ \bi{v'}_a = \bi{v}_a -
\hat{\bi{e}}_{ab}(\hat{\bi{e}}_{ab}\cdot\bi{v}_{ab})$ 
and $\bi{v'}_b =\bi{v}_b +
\hat{\bi{e}}_{ab}(\hat{\bi{e}}_{ab}\cdot\bi{v}_{ab})$.
Substituting equation (\ref{invple}) into this expression, we have the
contributions 
\begin{eqnarray}
  \fl
  4\rhom^2\frac{m^N}{(8\pi)^N|\mathcal{L}(N)|}
   \int \prod_{i=1}^N \ud \bi{r}_i \ud \bi{v}_i 
  \delta\left(\epsilon_i - \frac{m}{2} v_i^2\right)
  \int_{\hat{\bi{e}}_{ab}\cdot\bi{v}_{ab}>0}
  \ud\hat{\bi{e}}_{ab}(\hat{\bi{e}}_{ab}\cdot\bi{v}_{ab})
  \label{meqcomp2}
  \\
  \lo
  \times\Bigg[ \delta(\bi{r}_{ab} - 2\rhom \hat{\bi{e}}_{ab})
    \int \prod_{i=1}^N \ud \tilde\epsilon_i \tilde\epsilon_i^{-1}
  \delta\left(v'_i - \sqrt{\frac{2\tilde\epsilon_i}{m}}\right)
  P_N(\tilde\epsilon_1, \dots, \tilde\epsilon_N, t)
  \nonumber\\
  - \delta(\bi{r}_{ab} + 2\rhom\hat{\bi{e}}_{ab})
  \int \prod_{i=1}^N \ud \tilde\epsilon_i \tilde\epsilon_i^{-1}
  \delta\left( v_i - \sqrt{\frac{2\tilde\epsilon_i}{m}}\right)
  P_N(\tilde\epsilon_1, \dots, \tilde\epsilon_N, t)
  \Bigg]\,,
\nonumber
\end{eqnarray}
where $v'_i = v_i$ for $i\neq a,b$ and 
\begin{equation}
  \eqalign{
    v'_a = \sqrt{v_a^2 - (\widehat{\bi{e}}_{ab} \cdot \bi{v}_{a})^2
      + (\widehat{\bi{e}}_{ab} \cdot \bi{v}_{b})^2}\,,\\
    v'_b = \sqrt{v_b^2 + (\widehat{\bi{e}}_{ab} \cdot \bi{v}_{a})^2
      - (\widehat{\bi{e}}_{ab} \cdot \bi{v}_{b})^2}\,.
  }
\end{equation}
Inserting factors $\int \ud \eta \delta
(\eta \pm m/2 \big[(\widehat{\bi{e}}_{ab} \cdot \bi{v}_{a})^2
  - (\widehat{\bi{e}}_{ab} \cdot \bi{v}_{b})^2\big])$ in both of the terms
  between brackets, we write $\epsilon'_i = \epsilon_i$, $i\neq a,b$, and
  $\epsilon'_a =   \epsilon_a + \eta$, $\epsilon'_b =   \epsilon_b -
  \eta$. The contribution (\ref{meqcomp2}) thus transforms into
\begin{eqnarray}
  \fl
  4\rhom^2\frac{m^N}{(8\pi)^N|\mathcal{L}(N)|}
  \left(\frac{m}{2}\right)^{\frac{N}{2}}
   \int \prod_{i=1}^N \ud \bi{r}_i \ud \bi{v}_i 
  \delta\left(\epsilon_i - \frac{m}{2} v_i^2\right)
  \int_{\hat{\bi{e}}_{ab}\cdot\bi{v}_{ab}>0}
  \ud\hat{\bi{e}}_{ab}(\hat{\bi{e}}_{ab}\cdot\bi{v}_{ab})
  \nonumber  \\
  \lo
  \times\Bigg[ \delta(\bi{r}_{ab} - 2\rhom \hat{\bi{e}}_{ab})
  \int \ud \eta \delta
  \left(\eta + \frac{m}{2} \big[(\widehat{\bi{e}}_{ab} \cdot \bi{v}_{a})^2
  - (\widehat{\bi{e}}_{ab} \cdot \bi{v}_{b})^2\big]\right)
  \nonumber\\
  \times\int \prod_{i=1}^N \ud \tilde\epsilon_i \tilde\epsilon_i^{-1}
  \delta\big(\sqrt{}\epsilon'_i - \sqrt{}\tilde\epsilon_i\big)
  P_N(\tilde\epsilon_1, \dots, \tilde\epsilon_N, t)
  \nonumber\\
  - \delta(\bi{r}_{ab} + 2\rhom\hat{\bi{e}}_{ab})
  \int \ud \eta \delta
  \left(\eta - \frac{m}{2} \big[(\widehat{\bi{e}}_{ab} \cdot \bi{v}_{a})^2
  - (\widehat{\bi{e}}_{ab} \cdot \bi{v}_{b})^2\big]\right)
  \nonumber\\
  \times\int \prod_{i=1}^N \ud \tilde\epsilon_i \tilde\epsilon_i^{-1}
  \delta\big(\sqrt{}\epsilon_i - \sqrt{}\tilde\epsilon_i\big)
  P_N(\tilde\epsilon_1, \dots, \tilde\epsilon_N, t)
  \Bigg]\,,\nonumber\\
  \fl
  =   4\rhom^2\frac{m^N}{(8\pi)^N|\mathcal{L}(N)|}
  (2m)^{\frac{N}{2}}
   \int \prod_{i=1}^N \ud \bi{r}_i \ud \bi{v}_i 
  \delta\left(\epsilon_i - \frac{m}{2} v_i^2\right)
  \int_{\hat{\bi{e}}_{ab}\cdot\bi{v}_{ab}>0}
  \ud\hat{\bi{e}}_{ab}(\hat{\bi{e}}_{ab}\cdot\bi{v}_{ab})
  \nonumber  \\
  \lo
  \times\Bigg[ \delta(\bi{r}_{ab} - 2\rhom \hat{\bi{e}}_{ab})
  \int \ud \eta \delta
  \left(\eta + \frac{m}{2} \big[(\widehat{\bi{e}}_{ab} \cdot \bi{v}_{a})^2
  - (\widehat{\bi{e}}_{ab} \cdot \bi{v}_{b})^2\big]\right)
  \nonumber\\
  \times\frac{1}{\sqrt{\epsilon'_1 \dots\epsilon'_N}}
  P_N(\epsilon'_1, \dots,\epsilon'_N, t)
  \nonumber\\
  - \delta(\bi{r}_{ab} + 2\rhom\hat{\bi{e}}_{ab})
  \int \ud \eta \delta
  \left(\eta - \frac{m}{2} \big[(\widehat{\bi{e}}_{ab} \cdot \bi{v}_{a})^2
  - (\widehat{\bi{e}}_{ab} \cdot \bi{v}_{b})^2\big]\right)
  \nonumber\\
  \times\frac{1}{\sqrt{\epsilon_1 \dots\epsilon_N}}
  P_N(\epsilon_1, \dots,\epsilon_N, t)
  \Bigg]
  \label{meqcomp3}\,,
\end{eqnarray}
where we used the identity $\ud\epsilon/\epsilon =
2\ud\sqrt{\epsilon}/\sqrt{\epsilon}$ and carried out the $\tilde\epsilon$
integrations. We proceed by performing all the $\bi{r}_i$ and $\bi{v}_i$
integrations but for $i = a,b$. For the latter, we change variables from
$(\bi{r}_a, \bi{r}_b)$ to $(\bi{R}_{ab}, \bi{r}_{ab})$, the center of mass and
relative coordinates respectively. The latter integration can be carried
out, with outcome $\bi{r}_{ab} = \pm 2\rhom \hat{\bi{e}}_{ab}$. We are thus
led to contributions
\begin{eqnarray}
  \fl
  \frac{\rhom^2 m^3}{8\pi^2|\mathcal{L}(2)|}
   \int \ud \bi{R}_{ab} \ud \bi{v}_a  \ud \bi{v}_b
   \delta\left(\epsilon_a - \frac{m}{2} v_a^2\right)
   \delta\left(\epsilon_b - \frac{m}{2} v_b^2\right)
  \int_{\hat{\bi{e}}_{ab}\cdot\bi{v}_{ab}>0}
  \ud\hat{\bi{e}}_{ab}(\hat{\bi{e}}_{ab}\cdot\bi{v}_{ab})
  \nonumber  \\
  \times\Bigg[\frac{1}{\sqrt{(\epsilon_a + \eta)(\epsilon_b - \eta)}}
  \int \ud \eta \delta
  \left(\eta + \frac{m}{2} \big[(\widehat{\bi{e}}_{ab} \cdot \bi{v}_{a})^2
  - (\widehat{\bi{e}}_{ab} \cdot \bi{v}_{b})^2\big]\right)
  \nonumber\\
  \times P_N(\epsilon_1, \dots,\epsilon_a + \eta, \epsilon_b - \eta,
  \dots, \epsilon_N, t) 
  \nonumber\\
  - \frac{1}{\sqrt{\epsilon_a\epsilon_b }}
  \int \ud \eta \delta
  \left(\eta - \frac{m}{2} \big[(\widehat{\bi{e}}_{ab} \cdot \bi{v}_{a})^2
  - (\widehat{\bi{e}}_{ab} \cdot \bi{v}_{b})^2\big]\right)
  P_N(\epsilon_1, \dots,\epsilon_N, t)
  \Bigg]
  \label{meqcomp4}\,.
\end{eqnarray}
Notice that we have made the assumption that the position integrals of the
$N-2$ particles not involved in the collision factorise\footnote{All the
  cells are taken to be identical throughout.}:
\begin{equation}
  |\mathcal{L}(N)| =   |\mathcal{L}(2)| |\mathcal{L}(N-2)|\,.
\end{equation}
Though this is an approximation, it becomes exact under the assumption of
scale separation, which implies $|\mathcal{L}(N)| =
|\mathcal{L}(1)|^N$. Comparing the above expression to equation 
(\ref{mastereq}), we identify the kernels associated with the gain and loss
terms, 
\begin{eqnarray}
  \fl 
  W (\epsilon_a + \eta, \epsilon_b - \eta| \epsilon_a, \epsilon_b)
  =   \frac{\rhom^2m^3}{8\pi^2|\mathcal{L}(2)|}
  \int \ud\Omega\ud\bi{R}
  \nonumber\\   
  \times\frac{1}{\sqrt{(\epsilon_a + \eta) 
      (\epsilon_b - \eta)}}
  \int_{\widehat{\bi{e}}_{ab} \cdot \bi{v}_{ab}>0} 
  \ud\bi{v_a}\ud\bi{v_b} (\widehat{\bi{e}}_{ab} \cdot \bi{v}_{ab})
  \delta\left(\epsilon_a - \frac{m}{2} v_a^2\right) 
  \nonumber\\
  \times  
  \delta\left(\epsilon_b - \frac{m}{2} v_b^2\right) 
  \delta\left(\eta + \frac{m}{2}
    \big[(\widehat{\bi{e}}_{ab} \cdot \bi{v}_{a})^2
    - (\widehat{\bi{e}}_{ab} \cdot \bi{v}_{b})^2\big]\right)\,,
  \nonumber\\
  \lo
  =   -\frac{\rhom^2m^3}{8\pi^2|\mathcal{L}(2)|}
  \int \ud\Omega\ud\bi{R}
  \nonumber\\   
  \times\frac{1}{\sqrt{(\epsilon_a + \eta) 
      (\epsilon_b - \eta)}}
  \int_{\widehat{\bi{e}}_{ab} \cdot \bi{v'}_{ab}<0} 
  \ud\bi{v'_a}\ud\bi{v'_b} (\widehat{\bi{e}}_{ab} \cdot \bi{v'}_{ab})
  \delta\left(\epsilon_a + \eta - \frac{m}{2} {v'_a}^2\right) 
  \nonumber\\
  \times  
  \delta\left(\epsilon_b - \eta - \frac{m}{2} {v'_b}^2\right) 
  \delta\left(\eta + \frac{m}{2}
    \big[(\widehat{\bi{e}}_{ab} \cdot \bi{v'}_{b})^2
    - (\widehat{\bi{e}}_{ab} \cdot \bi{v'}_{a})^2\big]\right)\,,
  \label{kernelgain}\\
  \fl 
  W (\epsilon_a, \epsilon_b|\epsilon_a - \eta, \epsilon_b + \eta)
  =   \frac{\rhom^2m^3}{8\pi^2|\mathcal{L}(2)|}
  \int \ud\Omega\ud\bi{R}
  \nonumber\\   
  \times\frac{1}{\sqrt{\epsilon_a \epsilon_b}}
  \int_{\widehat{\bi{e}}_{ab} \cdot \bi{v}_{ab}>0} 
  \ud\bi{v_a}\ud\bi{v_b} (\widehat{\bi{e}}_{ab} \cdot \bi{v}_{ab})
  \delta\left(\epsilon_a - \frac{m}{2} v_a^2\right) 
  \nonumber\\
  \times  
  \delta\left(\epsilon_b - \frac{m}{2} v_b^2\right) 
  \delta\left(\eta - \frac{m}{2}
    \big[(\widehat{\bi{e}}_{ab} \cdot \bi{v}_{a})^2
    - (\widehat{\bi{e}}_{ab} \cdot \bi{v}_{b})^2\big]\right)\,,
  \label{kernelloss}
\end{eqnarray}
where, in the expression of the gain term (\ref{kernelgain}), we
changed the post-collisional velocity variables $(\bi{v}_a, \bi{v}_b)$ to
the pre-collisional velocities $(\bi{v'}_a, \bi{v'}_b)$.

Without loss of generality, we assume $\bi{v}_a$ and $\bi{v}_b$ are
measured with respect to referentials whose $z$-axis point along the unit
vector $\widehat{\bi{e}}_{ab}$. In such a case, we have
$\widehat{\bi{e}}_{ab} \cdot \bi{v}_{a} = v_a \cos\phi_a$ and
$\widehat{\bi{e}}_{ab} \cdot \bi{v}_{b} = v_b \cos\phi_b$. 
Thus the volume
integral $\int \ud\Omega\ud\bi{R}$ decouples from the
velocity integrals which yield
\begin{eqnarray}
  \fl
  - \int_{\widehat{\bi{e}}_{ab} \cdot \bi{v'}_{ab}<0} 
  \ud\bi{v'_a}\ud\bi{v'_b} (\widehat{\bi{e}}_{ab} \cdot \bi{v'}_{ab})
  \delta\left(\epsilon_a + \eta - \frac{m}{2} {v'_a}^2\right) 
  \delta\left(\epsilon_b - \eta - \frac{m}{2} {v'_b}^2\right) 
  \label{kernelgain1}\\
  \times\delta\left(\eta + \frac{m}{2}
    \big[(\widehat{\bi{e}}_{ab} \cdot \bi{v'}_{b})^2
    - (\widehat{\bi{e}}_{ab} \cdot \bi{v'}_{a})^2\big]\right)
  \nonumber\\
  \lo =   
  \pi^2\left(\frac{2}{m}\right)^{\frac{7}{2}}
  \sqrt{\epsilon_a'\epsilon'_b}
  \int_{\sqrt{\epsilon'_b}x_b > \sqrt{\epsilon'_a}x_a}
  \ud x_a \ud x_b 
  (\sqrt{\epsilon'_b}x_b - \sqrt{\epsilon'_a}x_a)
  \delta(\eta + [\epsilon'_b x_b^2 - \epsilon'_a x_a^2])\,,
  \nonumber\\
  \fl
  \int_{\widehat{\bi{e}}_{ab} \cdot \bi{v}_{ab}>0} 
  \ud\bi{v_a}\ud\bi{v_b} (\widehat{\bi{e}}_{ab} \cdot \bi{v}_{ab})
  \delta\left(\epsilon_a - \frac{m}{2} v_a^2\right) 
  \delta\left(\epsilon_b - \frac{m}{2} v_b^2\right) 
  \label{kernelloss1}\\
  \times\delta\left(\eta - \frac{m}{2}
    \big[(\widehat{\bi{e}}_{ab} \cdot \bi{v}_{a})^2
    - (\widehat{\bi{e}}_{ab} \cdot \bi{v}_{b})^2\big]\right)
  \nonumber\\
  \lo =   
  \pi^2\left(\frac{2}{m}\right)^{\frac{7}{2}}\sqrt{\epsilon_a \epsilon_b}
  \int_{\sqrt{\epsilon_a}x_a > \sqrt{\epsilon_b}x_b}
  \ud x_a \ud x_b 
  (\sqrt{\epsilon_a}x_a - \sqrt{\epsilon_b}x_b)
  \delta(\eta - [\epsilon_a x_a^2 - \epsilon_b x_b^2])\,.
  \nonumber
\end{eqnarray}
Notice that the two integrals are identical upon exchanging the roles of $a$
and $b$ and $\eta\to-\eta$, except for the energies $\epsilon_a' =
\epsilon_a + \eta$ and $\epsilon_b' = \epsilon_b - \eta$  in the expression
(\ref{kernelgain1}). Thus the kernels (\ref{kernelgain})-(\ref{kernelloss})
have expressions  
\begin{eqnarray}
  \fl 
  W (\epsilon_a + \eta, \epsilon_b - \eta| \epsilon_a, \epsilon_b)
  =   \sqrt{\frac{2}{m}}\frac{\rhom^2}{|\mathcal{L}(2)|}
  \int \ud\Omega\ud\bi{R}
  \nonumber\\   
  \times
  \int_{\sqrt{\epsilon'_b}x_b > \sqrt{\epsilon'_a}x_a}
  \ud x_a \ud x_b 
  (\sqrt{\epsilon'_b}x_b - \sqrt{\epsilon'_a}x_a)
  \delta(\eta + [\epsilon'_b x_b^2 - \epsilon'_a x_a^2])\,,
  \label{kernelgainc2}\\
  \fl 
  W (\epsilon_a, \epsilon_b|\epsilon_a - \eta, \epsilon_b + \eta)
  = \sqrt{\frac{2}{m}}\frac{\rhom^2}{|\mathcal{L}(2)|}
  \int \ud\Omega\ud\bi{R}
  \nonumber\\   
  \times
  \int_{\sqrt{\epsilon_a}x_a > \sqrt{\epsilon_b}x_b}
  \ud x_a \ud x_b 
  (\sqrt{\epsilon_a}x_a - \sqrt{\epsilon_b}x_b)
  \delta(\eta - [\epsilon_a x_a^2 - \epsilon_b x_b^2])\,.
  \label{kernellossc2}
\end{eqnarray}
Notice the symmetry between the kernels of the gain and loss terms, 
\emph{i.e.}
\begin{equation}
  W (\epsilon_a + \eta, \epsilon_b - \eta| \epsilon_a, \epsilon_b)
  = W (\epsilon'_a, \epsilon'_b|\epsilon'_a - \eta, \epsilon'_b + \eta)\,.
\end{equation}
This is however no to say that the kernels are symmetric. In fact
\begin{equation}
  W (\epsilon_a + \eta, \epsilon_b - \eta| \epsilon_a, \epsilon_b)
  \ne W (\epsilon_a, \epsilon_b|\epsilon_a + \eta, \epsilon_b - \eta)\,,
  \label{Wnotsym}
\end{equation}
which can be seen by rewriting equation (\ref{kernelgainc2}) directly in
terms of $\epsilon_a$ and $\epsilon_b$~:
\begin{eqnarray}
  \fl 
  W (\epsilon_a + \eta, \epsilon_b - \eta| \epsilon_a, \epsilon_b)
  =   \sqrt{\frac{2}{m}}\frac{\rhom^2}{|\mathcal{L}(2)|}
  \int \ud\Omega\ud\bi{R}
  \sqrt{\frac{\epsilon_a \epsilon_b}{(\epsilon_a + \eta) 
      (\epsilon_b - \eta)}}
  \nonumber\\   
  \times
  \int_{\sqrt{\epsilon_a}x_a > \sqrt{\epsilon_b}x_b}
  \ud x_a \ud x_b 
  (\sqrt{\epsilon_a}x_a - \sqrt{\epsilon_b}x_b)
  \delta(\eta + [\epsilon_a x_a^2 - \epsilon_b x_b^2])\,.
  \label{kernelgainc3}
\end{eqnarray}
The symmetry is therefore the following
\begin{equation}
  \sqrt{(\epsilon_a + \eta)(\epsilon_b - \eta)}
  W (\epsilon_a + \eta, \epsilon_b - \eta| \epsilon_a, \epsilon_b)
  = \sqrt{\epsilon_a \epsilon_b}
  W (\epsilon_a, \epsilon_b|\epsilon_a + \eta, \epsilon_b - \eta)\,,
\end{equation}
consistent with the form of the equilibrium distributions
(\ref{eqplecan}). This is to say that, despite the lack of symmetry
(\ref{Wnotsym}), detailed balance is recovered: 
\begin{eqnarray}
  \lo
  W (\epsilon_a + \eta, \epsilon_b - \eta| \epsilon_a, \epsilon_b)
  P^{(\mathrm{eq})}(\dots, \epsilon_a + \eta, \epsilon_b - \eta, \dots)
  \label{kernelsym}
  \\
  =   W (\epsilon_a, \epsilon_b|\epsilon_a + \eta, \epsilon_b - \eta)
  P^{(\mathrm{eq})}(\dots, \epsilon_a, \epsilon_b, \dots)\,.
  \nonumber
\end{eqnarray}
Therefore  $\partial_t P^{(\mathrm{eq})}(\epsilon_1,\dots \epsilon_N) =
0$, as must be.

\subsection{Computation of the loss term \label{sec.lossterm}}

Going back to equation (\ref{kernellossc2}), we can write
\begin{eqnarray}
  \lo
  \delta(\eta - \epsilon_a x_a^2 + \epsilon_b x_b^2)
  \label{deltaxa}\\
  = \frac{1}{2\sqrt{\epsilon_a(\eta + \epsilon_b x_b^2)}}
  \left[\delta\left(x_a - \sqrt{\frac{\eta + \epsilon_b x_b^2}{\epsilon_a}}
    \right) + \delta\left(x_a + \sqrt{\frac{\eta + \epsilon_b
          x_b^2}{\epsilon_a}}\right) \right]\,.
  \nonumber
\end{eqnarray}
The validity of this expression requires $\eta + \epsilon_b x_b^2 > 0$,
which is satisfied whenever $\eta>0$ or 
\begin{equation}
  \mathrm{abs}(x_b) >
  \sqrt{\frac{-\eta}{\epsilon_b}}\,,\quad\mathrm{if}\,\eta<0\,.
  \label{cond1}
\end{equation}
Provided this condition is fulfilled, we can carry out the
$x_a$-integration in equation (\ref{kernellossc2}). Considering the arguments of
the delta functions in equation (\ref{deltaxa}), the result of the
integration would be trivial unless
\begin{equation}
  \eta + \epsilon_b x_b^2 < \epsilon_a\Leftrightarrow
  \mathrm{abs}(x_b) < \sqrt{\frac{\epsilon_a - \eta}{\epsilon_b}}
  \,.
  \label{cond2}
\end{equation}

The bounds of the $x_b$-integral are respectively
\begin{equation}
  \eqalign{
    x_\mathrm{min} \equiv \mathrm{max}[-1,-\sqrt{(\epsilon_a - \eta) 
      /\epsilon_b}],\\
    x_\mathrm{max} \equiv \mathrm{min}[1,\sqrt{(\epsilon_a - \eta)
      /\epsilon_b}].
    }
\end{equation}
which is
\begin{equation}
  x_\mathrm{max} = -  x_\mathrm{min} =
  \left\{
    \begin{array}{l@{\quad}l}
      1\,,
      & \eta<\epsilon_a - \epsilon_b\,,\\
      \sqrt{\frac{\epsilon_a - \eta}{\epsilon_b}}\,,
      &\eta > \epsilon_a - \epsilon_b\,.
    \end{array}
  \right.
 \label{xmax}
\end{equation}

\subsubsection{$\eta>0$:}

The condition on the integration domain transposes
into the Heaviside step function with 
arguments $\pm\sqrt{\eta + \epsilon_b x_b^2} - \sqrt{\epsilon_b} x_b$. 
Only the positive sign contributes to the kernel
(\ref{kernellossc2}). Leaving aside the prefactors, we have
\begin{eqnarray}
  W(\epsilon_a,\epsilon_b|\epsilon_a-\eta,\epsilon_b+\eta) &\sim
  \frac{1}{2 \sqrt{\epsilon_a}} 
  \int_{x_\mathrm{min}}^{x_\mathrm{max}} \ud x_b 
  \left(1 - \frac{\sqrt{\epsilon_b}
      x_b}{\sqrt{\eta + \epsilon_b x_b^2}}\right)\,,
    \nonumber\\
    &= \frac{1}{\sqrt{\epsilon_a}} x_\mathrm{max}\,,
\end{eqnarray}
where we used the symmetry $x_\mathrm{min} = -x_\mathrm{max}$. 
Using equation (\ref{xmax}), we have
\begin{equation}
  \lo
  W(\epsilon_a,\epsilon_b|\epsilon_a-\eta,\epsilon_b+\eta) \sim
  \left\{
    \begin{array}{l@{\quad}l}
      \frac{1}{\sqrt{\epsilon_a}}\,,          
      & 0<\eta<\epsilon_a - \epsilon_b\quad
      (\mathrm{if}\,\epsilon_a > \epsilon_b)\,,\\
      \sqrt{\frac{\epsilon_a - \eta}{\epsilon_a\epsilon_b}}\,,
      &0\le \epsilon_a - \epsilon_b < \eta < \epsilon_a\,,
    \end{array}
  \right.
\end{equation}

\subsubsection{$\eta<0$:}

In this case, the $x_b$-integral in (\ref{kernellossc2}) splits into two
separate integrals, the 
first from $x_\mathrm{min}$, as defined above, to $-x_\mathrm{mid}$,
and the second from $x_\mathrm{mid}$ to $x_\mathrm{max}$, with
\begin{equation}
  x_\mathrm{mid} \equiv \sqrt{\frac{-\eta}{\epsilon_b}}\,.
\end{equation}
Note that the bounds are well ordered~: we always have $x_\mathrm{mid} \le
x_\mathrm{max}$. 

The terms contributing to the kernel are
$\pm\sqrt{\eta + \epsilon_b x_b^2} - \sqrt{\epsilon_b} x_b$ for 
$x_b<0$. We have
\begin{eqnarray}
  \fl 
  W(\epsilon_a,\epsilon_b|\epsilon_a-\eta,\epsilon_b+\eta) &\sim
  \frac{1}{2\sqrt{\epsilon_a}}
  \int_{-x_\mathrm{max}}^{-x_\mathrm{mid}} \ud x_b 
  \left(- 2\frac{\sqrt{\epsilon_b}
      x_b}{\sqrt{\eta + \epsilon_b x_b^2}}\right)\,,
    \nonumber\\
    &= \sqrt{\frac{\eta + \epsilon_b x_\mathrm{max}^2}
      {\epsilon_a\epsilon_b}} - 
    \sqrt{\frac{\eta + \epsilon_b x_\mathrm{mid}^2}      
      {\epsilon_a\epsilon_b}}
\end{eqnarray}
Substituting the expressions of $x_\mathrm{max}$ and $x_\mathrm{mid}$, we
find
\begin{equation}
  W(\epsilon_a,\epsilon_b|\epsilon_a-\eta,\epsilon_b+\eta) \sim
  \left\{
    \begin{array}{l@{\quad}l}
      \frac{1}{\sqrt{\epsilon_b}}\,,
      & \epsilon_a - \epsilon_b<\eta<0\quad
      (\mathrm{if}\,\epsilon_b > \epsilon_a)\,,\\
      \sqrt{\frac{\eta + \epsilon_b}{\epsilon_a\epsilon_b}}\,,
      &- \epsilon_b < \eta < \epsilon_a - \epsilon_b\le0\,.
    \end{array}
  \right.
\end{equation}

\subsubsection{To summarize:}

We can write for the loss term:
\begin{eqnarray}
  W(\epsilon_a,\epsilon_b|\epsilon_a-\eta,\epsilon_b+\eta) = 
  \sqrt{\frac{2}{m}}\frac{\rhom^2}{|\mathcal{L}(2)|}
  \int \ud\Omega\ud\bi{R}
  \label{kernelfin}\\
  \hspace{2cm} \times\left\{
    \begin{array}{l@{\quad}l}
      \sqrt{\frac{\epsilon_a - \eta}{\epsilon_a \epsilon_b}}\,,& 
      \mathrm{max}(0,\epsilon_a - \epsilon_b) < \eta <
      \epsilon_a\,,\\ 
      \frac{1}{\sqrt{\mathrm{max}(\epsilon_a, \epsilon_b)}}\,,
      & \mathrm{min}(0,\epsilon_a - \epsilon_b)
      < \eta <  \mathrm{max}(0,\epsilon_a - \epsilon_b)\,,\\
      \sqrt{\frac{\epsilon_b + \eta}{\epsilon_a\epsilon_b}}
      \,,& -\epsilon_b < \eta <  \mathrm{min}(0,\epsilon_a -
      \epsilon_b)\,.
    \end{array}
  \right.
  \nonumber
\end{eqnarray}
It is readily checked that $W$ is symmetric under exchanging 
$\epsilon_a \leftrightarrow \epsilon_b$ and $\eta\to-\eta$,
as it must be, and has the inverse units of time times energy.

\subsection{Gain term}

Going through the same steps as in Section \ref{sec.lossterm} above, we
find for the gain term: 
\begin{eqnarray}
  W(\epsilon_a + \eta,\epsilon_b -\eta|\epsilon_a,\epsilon_b) = 
  \sqrt{\frac{2}{m}}\frac{\rhom^2}{|\mathcal{L}(2)|}
  \int \ud\Omega\ud\bi{R}
  \label{kernelg}\\
  \hspace{2cm} \times\left\{
    \begin{array}{l@{\quad}l}
      \frac{1}{\sqrt{\epsilon_a + \eta}}\,,& 
      \mathrm{max}(\epsilon_b - \epsilon_a,0) < \eta < \epsilon_b\,,\\ 
      \sqrt{\frac{\mathrm{min}(\epsilon_a,\epsilon_b)}
        {(\epsilon_a + \eta)(\epsilon_b - \eta)}}\,,
      & \mathrm{min}(\epsilon_b - \epsilon_a,0)  <  \eta < 
      \mathrm{max}(\epsilon_b - \epsilon_a,0)\,,\\
      \frac{1}{\sqrt{\epsilon_b - \eta}}\,,& 
      -\epsilon_a < \eta < \mathrm{min}(\epsilon_b - \epsilon_a,0) \,.
    \end{array}
  \right.
  \nonumber
\end{eqnarray}

Comparing with equation (\ref{kernelfin}), we check that
the loss and gain terms have the symmetry of equation (\ref{kernelsym}).

\section{Collision frequency and heat conductivity \label{sec.kappa}}

\begin{figure}[htb]
  \centering
  \includegraphics[width=.7\textwidth]{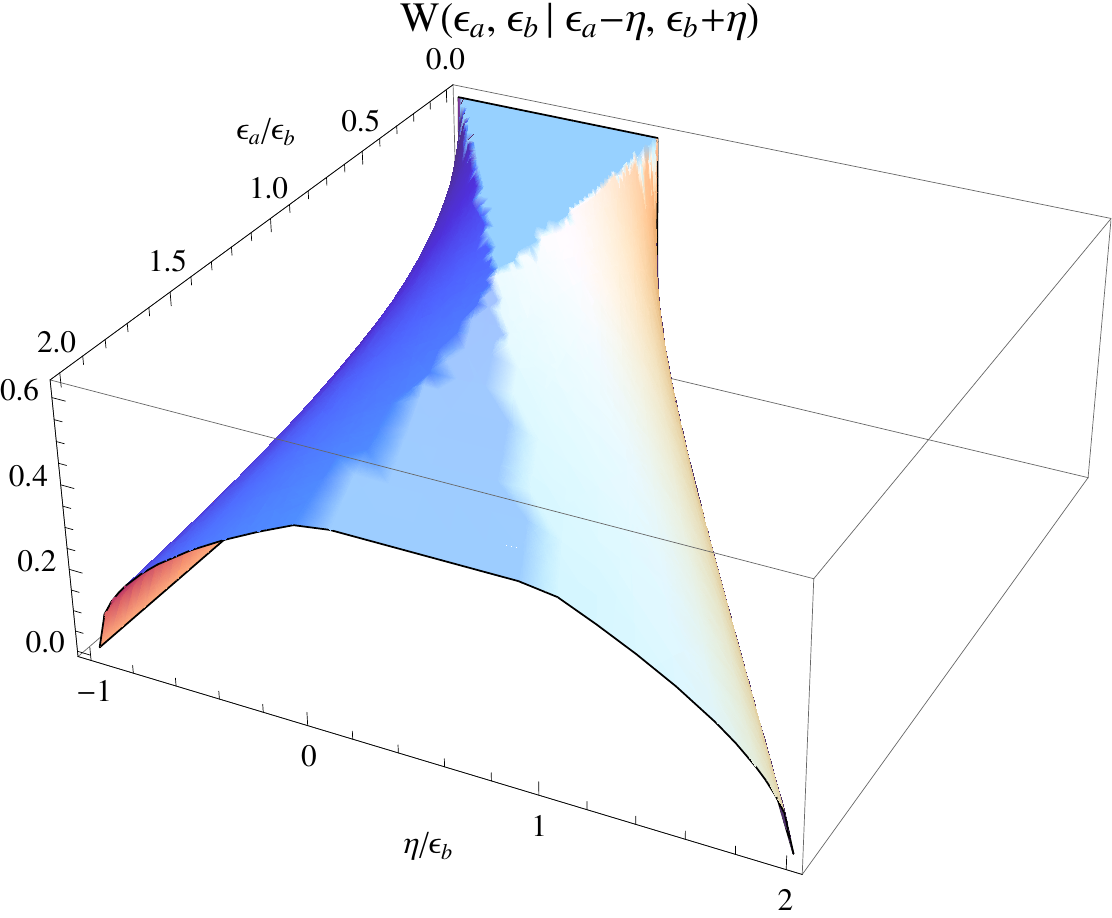}
  \caption{Stochastic kernel (\ref{kernelrsc}) associated with the
    probability of an energy exchange $\eta$ between two cells at energies
    $\epsilon_a$ and $\epsilon_b$.} 
  \label{fig.kernel}
\end{figure}

We can rescale time by a factor
$(4\rhom)^2/(\sqrt{m\pi}|\mathcal{L}(2)|) \int \ud\Omega\ud\bi{R}$, which
amounts to converting the time units to units of the inverse square root of
energy. The expression of the kernel thus simplifies to:
\begin{eqnarray}
    W(\epsilon_a,\epsilon_b|\epsilon_a-\eta,\epsilon_b+\eta) = 
    \label{kernelrsc}\\
    \hspace{2cm} \sqrt{\frac{\pi}{8}}\times\left\{
      \begin{array}{l@{\quad}l}
        \sqrt{\frac{\epsilon_b + \eta}{\epsilon_a\epsilon_b}}
        \,,& -\epsilon_b < \eta <  \mathrm{min}(0,\epsilon_a -
        \epsilon_b)\,,\\
        \frac{1}{\sqrt{\mathrm{max}(\epsilon_a, \epsilon_b)}}\,,
        & \mathrm{min}(0,\epsilon_a - \epsilon_b)
        < \eta <  \mathrm{max}(0,\epsilon_a - \epsilon_b)\,,\\
        \sqrt{\frac{\epsilon_a - \eta}{\epsilon_a \epsilon_b}}\,,& 
        \mathrm{max}(0,\epsilon_a - \epsilon_b) < \eta <
        \epsilon_a\,.
      \end{array}
    \right.
    \nonumber
\end{eqnarray}
A graphical representation of this function is displayed in figure
\ref{fig.kernel}. 

The rate of energy exchange (\ref{collfreq}) is
\begin{eqnarray}
  \nu(\epsilon_a, \epsilon_b) 
  =\frac{\sqrt{2\pi}}{12}\times \left\{
    \begin{array}{l@{\quad}l}
      \frac{\epsilon_a + 3\epsilon_b}{\sqrt{\epsilon_b}}\,,& 
      \epsilon_b \ge \epsilon_a\,,\\
      \frac{3\epsilon_a + \epsilon_b}{\sqrt{\epsilon_a}}\,,& 
      \epsilon_a \ge \epsilon_b\,,\\
    \end{array}
  \right.
\end{eqnarray}
with equilibrium average taken with respect to the measure (\ref{eqplecan}),
\begin{equation}
  \nu_\mathrm{B} = \frac{4}{\pi T^3}\int \ud\epsilon_a \ud \epsilon_b
  \nu(\epsilon_a, \epsilon_b) \sqrt{\epsilon_a \epsilon_b}
  e^{-(\epsilon_a + \epsilon_b)/T}
  = \sqrt{T}\,.
  \label{nub}
\end{equation}
The micro-canonical average of the collision frequency is the quantity
\begin{equation}
  \nu_\mathrm{B}(N) = \int \ud\epsilon_a \ud \epsilon_b
  \nu(\epsilon_a, \epsilon_b) P^{(\mathrm{mic})}_N(\epsilon_a, \epsilon_b)
  = \sqrt{T}[1 + \mathcal{O}(1/N)]\,.
  \label{nubmic}
\end{equation}

The amount of heat transfer is
\begin{eqnarray}
  j(\epsilon_a, \epsilon_b) 
  = \frac{\sqrt{2\pi}}{24}(\epsilon_a - \epsilon_b)
  \times \left\{
    \begin{array}{l@{\quad}l}
      \frac{\epsilon_a +
        3\epsilon_b}{\sqrt{\epsilon_b}}\,, 
      & \epsilon_b \ge \epsilon_a\,,\\
      \frac{3 \epsilon_a + \epsilon_b}{\sqrt{\epsilon_a}}\,,
      & \epsilon_a \ge \epsilon_b\,,\\
    \end{array}
  \right.
\end{eqnarray}
which is equivalent to 
\begin{equation}
  j(\epsilon_a, \epsilon_b) = \frac{1}{2}(\epsilon_a -
  \epsilon_b) \nu(\epsilon_a, \epsilon_b). 
\end{equation}
Thus, with the help of the identity
\begin{equation}
  \frac{1}{\pi T^5}
  \int \ud\epsilon_a \ud \epsilon_b
  \nu(\epsilon_a, \epsilon_b) (\epsilon_a - \epsilon_b)^2
  \sqrt{\epsilon_a \epsilon_b}
  e^{-(\epsilon_a + \epsilon_b)/T}
  =\sqrt{T}\,,
  \label{idnu}
\end{equation}
we can compute the average of the heat current weighted by $(\epsilon_a -
\epsilon_b)/2$ with respect to local thermal equilibrium measure to obtain
the heat conductivity:
\begin{eqnarray}
  \kappa &= \frac{2}{\pi T^5}\int \ud\epsilon_a \ud \epsilon_b
  j(\epsilon_a, \epsilon_b) (\epsilon_a - \epsilon_b) 
  \sqrt{\epsilon_a \epsilon_b}
  e^{-(\epsilon_a + \epsilon_b)/T}\,,\nonumber\\
  &=  \frac{1}{\pi T^5}\int \ud\epsilon_a \ud \epsilon_b
  \nu(\epsilon_a, \epsilon_b) (\epsilon_a - \epsilon_b)^2
  \sqrt{\epsilon_a \epsilon_b}
  e^{-(\epsilon_a + \epsilon_b)/T}\,,\nonumber\\
  &= \sqrt{T}\,,
\end{eqnarray}
The ratio between the heat conductivity and collision frequency 
is therefore unity, as announced (\ref{kappaeqnu}).

The same result is obtained from the equilibrium average of the second
moment of the heat transfer (\ref{etasq}):
\begin{equation}
  h(\epsilon_a, \epsilon_b) = 
  \frac{\sqrt{2\pi}}{420}
    \left\{
      \begin{array}{l@{\quad}l}
        \frac{35 \epsilon_b^2(\epsilon_b -  \epsilon_a) + 21  \epsilon_b
          \epsilon_a^2 + 11 \epsilon_a^3}{\sqrt{\epsilon_b}}\,,
        & \epsilon_b \ge  \epsilon_a\,,\\
        \frac{35 \epsilon_a^2(\epsilon_a -  \epsilon_b) + 21  \epsilon_a
          \epsilon_b^2 + 11 \epsilon_b^3}{\sqrt{\epsilon_a}}\,,
        & \epsilon_a \ge  \epsilon_b\,,
      \end{array}
    \right.
\end{equation}
which yields an alternative derivation of the heat conductivity
by taking the equilibrium average of this quantity:
\begin{eqnarray}
  \kappa &= \frac{2}{\pi T^5}\int \ud\epsilon_a \ud \epsilon_b
  h(\epsilon_a, \epsilon_b)
  \sqrt{\epsilon_a \epsilon_b}
  e^{-(\epsilon_a + \epsilon_b)/T}\,,\nonumber\\
  &= \sqrt{T}\,.
\end{eqnarray}

This result is universal for three-dimensional systems in the sense that it
holds for confined particles interacting through hardcore collisions,
independent of the shape of the confining cells and temperature, at least
so long as the local dynamics is ergodic on the constant energy surface. 

\section{Numerical scheme \label{sec.num}}

Equation (\ref{kappaeqnu}) can be verified by direct numerical computations
of the master equation (\ref{mastereq}) according to the scheme described
in \cite{GG08c}, based on Gillespie's algorithm \cite{Gil76,Gil77}.
We recall that the Monte-Carlo step necessitates two random trials. The
first random number determines the time that will elapse until the next
energy exchange event. The second one determines which one out of all the
possible pairs of cells will perform an exchange of energy and how much
energy will be exchanged between them. 

Thus let $\xi$ and $\chi$ be these two random numbers, uniformly
distributed on the unit interval and let us consider an array --whether
one-, two- or three-dimensional-- of $N$ \emph{pairs}\footnote{The
  corresponding number of energy variables depends on the dimensionality of
  the lattice. In dimension 1, each energy variable is involved in two
  pairs; in dimension 2, four pairs; in dimension 3, 6 pairs.} of energy
variables $\Big\{\epsilon_1^{(i)},\epsilon_2^{(i)}\Big\}_{i=1}^N$.

The first random number, $\xi$, determines the time to
the next energy exchange event, denoted $\tau$, according to the Poisson
distribution whose relaxation rate is given by the sum of the respective
exchange rates $\nu\Big(\epsilon_1^{(i)},\epsilon_2^{(i)}\Big)$ between
pairs of energies $\Big\{\epsilon_1^{(i)},\epsilon_2^{(i)}\Big\}$,
\begin{equation}
  \tau = \frac{\log \xi^{-1}}{\sum_{i=1}^N
    \nu\Big(\epsilon_1^{(i)},\epsilon_2^{(i)}\Big)} 
\end{equation}
The second random number, $\chi$, determines the pair $n$ of energies
$\Big(\epsilon_1^{(n)},\epsilon_2^{(n)}\Big)$ undergoing the energy
exchange and how much energy $\eta$ they exchange, according to 
\begin{equation}
  \fl
  \chi = \frac{1}{\sum_{i=1}^N
    \nu\Big(\epsilon_1^{(i)},\epsilon_2^{(i)}\Big)} \left[\sum_{i =
      0}^{n-1}  \nu\Big(\epsilon_1^{(i)},\epsilon_2^{(i)}\Big) 
    + \int_{-\epsilon_1^{(n)}}^\eta \ud \eta'
  W\Big(\epsilon_1^{(n)}, \epsilon_2^{(n)} | \epsilon_1^{(n)} - \eta,
  \epsilon_2^{(n)} + \eta\Big)\right]\,, 
\end{equation}
which must be solved for $\eta$. Thus, letting
\begin{equation}
   x \equiv \chi \sum_{i=1}^N
  \nu\Big(\epsilon_1^{(i)},\epsilon_2^{(i)}\Big) - \sum_{i = 0}^{n-1}
  \nu\Big(\epsilon_1^{(i)},\epsilon_2^{(i)}\Big)\,,
\end{equation}
and assuming 
\begin{equation}
  0 < x  - \int_{-\epsilon_1^{(n)}}^\eta \ud \eta'
  W\Big(\epsilon_1^{(n)}, \epsilon_2^{(n)}|\epsilon_1^{(n)} - \eta,
  \epsilon_2^{(n)} + \eta\Big) < \nu\Big(\epsilon_1^{(n)},
  \epsilon_2^{(n)}\Big)\,,
\end{equation}
we have
\begin{equation}
  \fl
  \eta = 
  \left\{
  \begin{array}{l@{\quad}l}
    3\left(\frac{2 x^2 \epsilon_1^{(n)}
        \epsilon_2^{(n)}}{3\pi}\right)^{1/3} 
    - \epsilon_2^{(n)}\,,&
    0 \leq x < \nu_1\Big(\epsilon_1^{(n)},\epsilon_2^{(n)}\Big)\,,\\
    2 x\sqrt{\frac{2}{\pi}} 
    \mathrm{max}\Big(\sqrt{\epsilon_1^{(n)}}, \sqrt{\epsilon_2^{(n)}}
    \Big)
      - \frac{2}{3} \mathrm{min}\Big(\epsilon_1^{(n)},\epsilon_2^{(n)}
      \Big)\\
      + \mathrm{min}\Big(0, \epsilon_1^{(n)}-\epsilon_2^{(n)}\Big)
    \,,&
    \nu_1\Big(\epsilon_1^{(n)},\epsilon_2^{(n)}\Big) \leq x < 
    \nu_2\Big(\epsilon_1^{(n)},\epsilon_2^{(n)}\Big)\,,\\
    \epsilon_1^{(n)} - 3\left(\frac{2 \epsilon_1^{(n)} 
        \epsilon_2^{(n)}}{3\pi}\right)^{1/3}
    \Big[\nu\Big(\epsilon_1^{(n)},\epsilon_2^{(n)}\Big) - x\Big]^{2/3}
    \,,&
    \nu_2\Big(\epsilon_1^{(n)},\epsilon_2^{(n)}\Big) \leq x < 
    \nu\Big(\epsilon_1^{(n)},\epsilon_2^{(n)}\Big)\,.
  \end{array}
  \right.
\end{equation}
In the above equation, we introduced the intermediate bounds $\nu_1$ and
$\nu_2$ which are computed through the partial integrals
\begin{eqnarray}
  \fl
  \int_{-\epsilon_b}^\eta \ud \eta'
    W(\epsilon_a,\epsilon_b|\epsilon_a-\eta',\epsilon_b+\eta') = 
    \label{kernelint}\\
    \hspace{-1cm} \sqrt{\frac{\pi}{2}}\times\left\{
      \begin{array}{l@{\quad}l}
        \frac{\left(\eta + \epsilon_b\right)^{\frac{3}{2}}}
        {3\sqrt{\epsilon_a \epsilon_b}}
        \,,& -\epsilon_b < \eta <  \mathrm{min}(0,\epsilon_a -
        \epsilon_b)\,,\\
        \frac{2\mathrm{min}(\epsilon_a, \epsilon_b)
        +
        3[\eta - \mathrm{min}(0, \epsilon_a - \epsilon_b)]}
        {6 \mathrm{max}(\sqrt{\epsilon_a}, \sqrt{\epsilon_b})}
        \,,
        & \mathrm{min}(0,\epsilon_a - \epsilon_b)
        < \eta < \\
        & \hspace{3cm}
        \mathrm{max}(0,\epsilon_a - \epsilon_b)\,,\\
        \frac{1}{3}\left[
          \frac{\epsilon_a + \epsilon_b + 
            2\mathrm{max}(\epsilon_a, \epsilon_b)}
          {2\mathrm{max}(\sqrt{\epsilon_a}, \sqrt{\epsilon_b})}
          - \frac{\left(\epsilon_a - \eta \right)^{\frac{3}{2}}}
          {\sqrt{\epsilon_a \epsilon _b}}\right]
        \,,& 
        \mathrm{max}(0,\epsilon_a - \epsilon_b) < \eta <
        \epsilon_a\,.
      \end{array}
    \right.
    \nonumber
\end{eqnarray}
In particular,
\begin{eqnarray}
  \nu_1(\epsilon_a, \epsilon_b) &\equiv
  \int_{-\epsilon_b}^{\mathrm{min}(0,\epsilon_a - \epsilon_b)} 
  \ud \eta
  W(\epsilon_a,\epsilon_b|\epsilon_a-\eta,\epsilon_b+\eta)\,,
  \nonumber\\
  &= 
  \frac{\sqrt{2\pi}}{6}\frac{\mathrm{min}(\epsilon_a,\epsilon_b)}
  {\mathrm{max}(\sqrt{\epsilon_a}, \sqrt{\epsilon_b})}
  \,,
  \label{nu1}\\
  \nu_2(\epsilon_a, \epsilon_b) &\equiv
  \int_{-\epsilon_b}^{\mathrm{max}(0,\epsilon_a - \epsilon_b)} 
  \ud \eta
  W(\epsilon_a,\epsilon_b|\epsilon_a-\eta,\epsilon_b+\eta)\,,
  \nonumber\\
  &= 
  \nu(\epsilon_a, \epsilon_b) - 
  \frac{\sqrt{2\pi}}{6}\frac{\mathrm{min}(\epsilon_a,\epsilon_b)}
  {\mathrm{max}(\sqrt{\epsilon_a}, \sqrt{\epsilon_b})}
  \,,\nonumber\\
  &=\nu(\epsilon_a, \epsilon_b) - \nu_1(\epsilon_a, \epsilon_b)\,.
  \label{nu2}
\end{eqnarray}

We remark that the scheme is explicit here and does not require using root
finding algorithms to determine the amount $\eta$ of energy exchanged, as
was the case with the kernel associated with two-dimensional underlying
dynamics in \cite{GG08c}. This considerably reduces the amount of CPU time
necessary to implement the algorithm.

\subsection{Equilibrium simulation}

We briefly discuss the results of equilibrium simulations performed along
this scheme on one- and two-dimensional arrays of $N$ cells\footnote{Notice
  the change of notation: $N$ refers to the system size here and not the
  number of pairs of neighbouring cells.} with periodic
boundary conditions, \emph{i.e.} identifying cells $N+1$ and $1$ in the
one-dimensional case and working with a square lattice of size $\sqrt{N}
\times \sqrt{N}$ and identifying the first and last columns and rows in the
two-dimensional case.

\begin{figure}[htb]
  \centering
  \includegraphics[width=.7\textwidth]{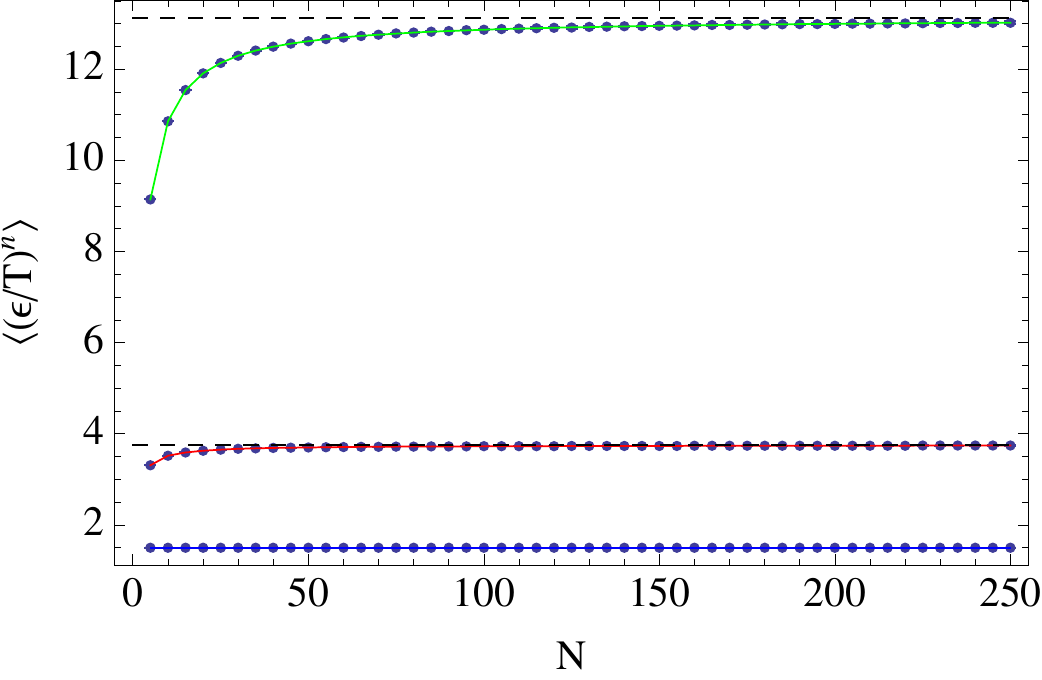}
  \vskip .5cm
  \includegraphics[width=.7\textwidth]{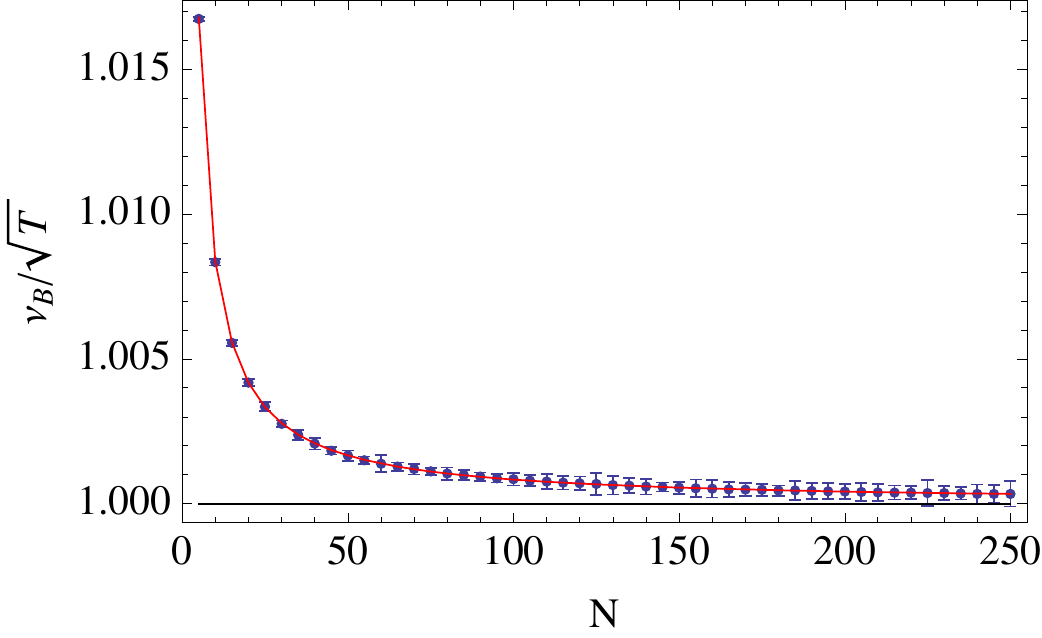}
  \caption{(Top) Energy moments, $\langle (\epsilon/T)^n \rangle$, for
    $n=1,2,3$, for values of $N$ up to $N=250$. The blue dots are the
    results of numerical computations of these quantities, using the
    scheme described above, and the solid lines correspond to their
    theoretical values derived in equation (\ref{micenfirst}). The dashed
    line correspond to the canonical expectations
    (\ref{canenmoment}). (Bottom) 
    Micro-canonical energy exchange frequency vs. $N$. The blue dots are
    the results of numerical computations of the energy exchange frequency
    for the corresponding value of $N$. The solid red line is the numerical
    integration of the equation (\ref{nubmic}).} 
  \label{fig.nubmic}
\end{figure}

The top panel of figure \ref{fig.nubmic} shows the results of a numerical 
computation of the first three energy moments for the one-dimensional
lattice with different values of $N$ up to $N=250$ and 
compares it to equation (\ref{micenfirst}). The accuracy of the calculation
demonstrates the invariance of the one-cell distributions
(\ref{micple1pt}), and by extension, of the micro-canonical distribution
(\ref{eqplemic}). This is further validated by a computation of the
energy exchange frequency, which involves the two-cell distribution
(\ref{micple2pt}). The bottom panel of figure \ref{fig.nubmic} thus 
shows the results of a numerical computation of this quantity for 
values of $N$ up to $N=250$, together with a comparison to a 
numerical evaluation of equation (\ref{nubmic}). Similar results are
obtained for two-dimensional lattices.
 
\begin{figure}[htbp]
  \centering
  \includegraphics[width=.7\textwidth]{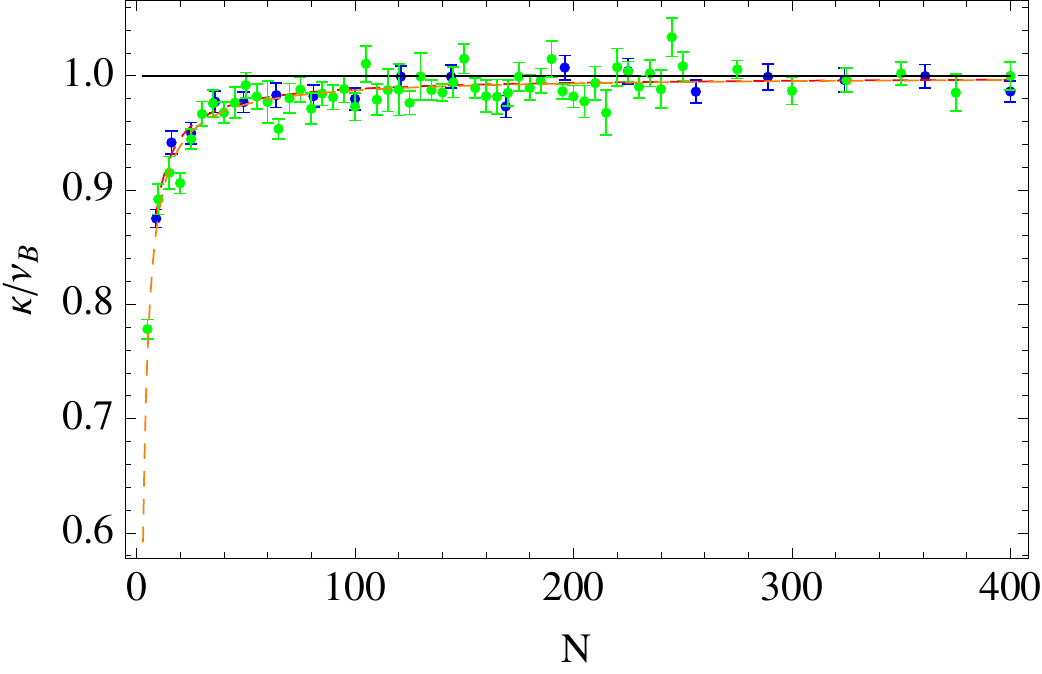}
  \caption{Ratio between the mean squared displacement of the Helfand
    moment and energy exchange frequency for one- (green) and
    two-dimensional (blue) lattices of cells vs. the size of the lattices
    $N$. The ratio $\kappa/\nu_\mathrm{B}$ is the  
    infinite $N$ extrapolation of these data, which we evaluate by
    linear regression (dashed red line). Fitting the data points 
    corresponding to $N\ge 9$ (of which
    only a fraction are displayed here)
    with weights inversely proportional to the
    sizes of their error bars, we obtain, for the one-dimensional lattice, 
    $\kappa/\nu_\mathrm{B} = 0.9994 \pm 0.0016$, with $N$ up to 400, and,
    for the two-dimensional lattice, $\kappa/\nu_\mathrm{B} = 0.9993 \pm
    0.0026$, with $N$ up to $400$ ($=20\times20$).} 
  \label{fig.kappa}
\end{figure}

The computation of the mean-squared displacement of the Helfand moment,
$H(t) = \sum_{a = 1}^N a \epsilon_a(t)$, yields the conductivity in the
limit of large system sizes,
\begin{equation}
  \kappa = \lim_{N\to\infty} \frac{1}{N (E/N)^2} 
  \lim_{n\to\infty} 
  \bigg \langle \frac{1}{2\tau_n}  \Delta H(\tau_n)^2 \bigg\rangle_{E/N}\,.
\label{kappahelf}
\end{equation}
As seen in figure \ref{fig.kappa}, the conductivity is identical to the
collision frequency, in accordance with equation (\ref{kappaeqnu}):
\begin{equation}
  \frac{\kappa}{\nu_\mathrm{B}} = 
  \left\{
    \begin{array}{l@{\quad}l}
      0.9994 \pm 0.0016\,,&
      (1D)\,,\\
      0.9993 \pm 0.0026\,,&
      (2D)\,.
    \end{array}
  \right.
\end{equation}

\subsection{Non-equilibrium simulation}

Non-equilibrium boundary conditions are easily implemented on
one-dimensional lattices by thermalising the boundary cells at respective
temperatures $T_- = 0.5$ and $T_+ = 1.5$ according to the scheme detailed
in \cite{GG08c}. 

The heat conductivity is evaluated by computing the stationary heat current, 
$\JH$, and comparing it to the temperature gradient according to 
\begin{equation}
  \JH = - \kappa \frac{T_+ - T_-}{N}\,,
\end{equation}
in the limit of large system sizes $N$.

Denoting by $\Ptwo_{a,a+1}(\epsilon, \epsilon')$ 
the two-point marginal of the stationary state associated with cells 
$a$ and $a+1$ at respective energies $\epsilon$ and $\epsilon'$, the heat 
current is
\begin{equation}
  \JH \equiv \int \ud \epsilon \ud \epsilon'
  j(\epsilon,\epsilon') \Ptwo_{a,a+1}(\epsilon, \epsilon')\,.
  \label{currenttwop}
\end{equation}

\begin{figure}[htbp]
  \centering
  \includegraphics[width=.5\textwidth]{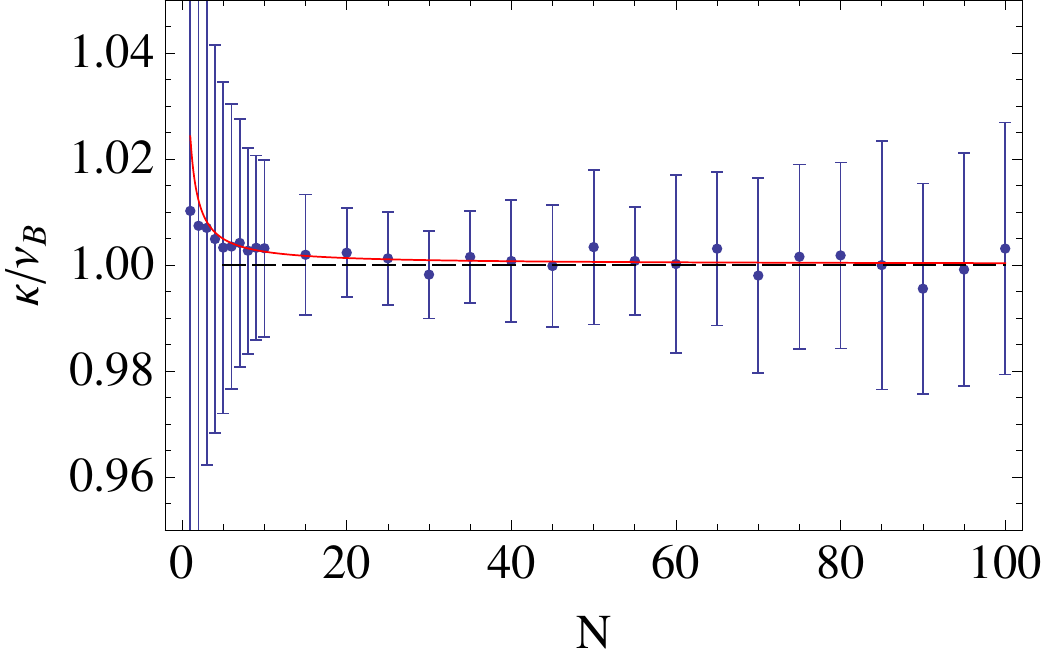}
  \includegraphics[width=.48\textwidth]{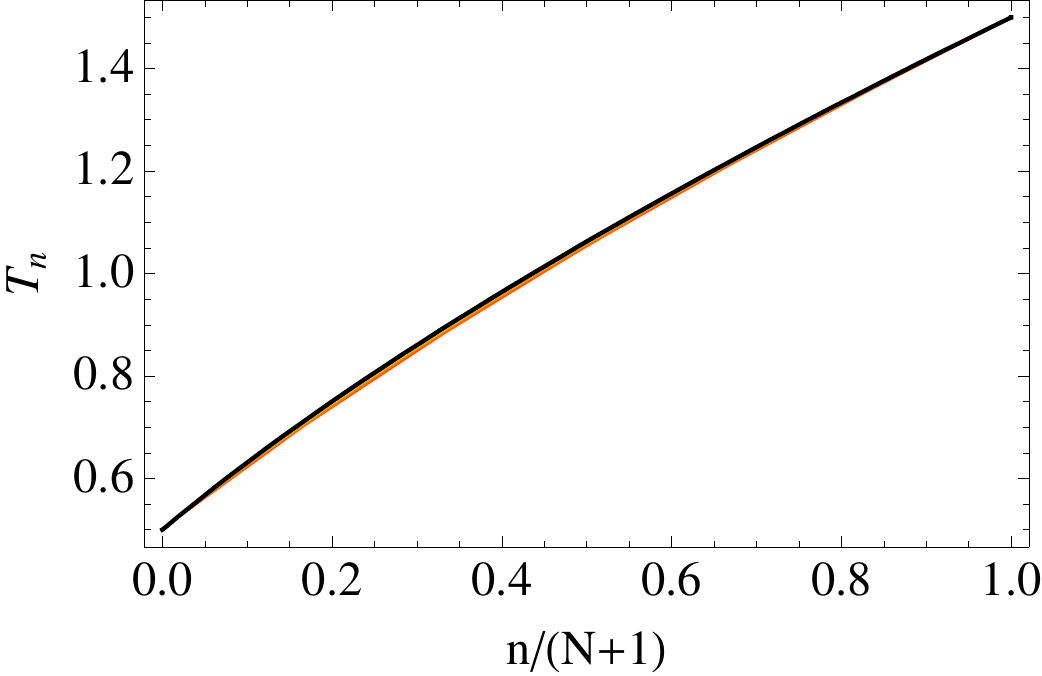}
  \caption{
    (Left) Ratio between the heat current (\ref{currenttwop}) and
    local temperature gradient under thermal boundary conditions at 
    the temperatures $T_+ = 1.5$ and $T_- = 0.5$, corresponding to the 
    temperature profiles shown above. The system sizes $N$ here range 
    from $N=1$ to $N=100$. The ratio $\kappa/\nub$ is computed from the 
    infinite $N$ extrapolation of these data, evaluated by linear 
    regression with respect to $1/N$ (solid red line). Fitting the data 
    points weighed according to the sizes of their error bars, we 
    obtain $\kappa/\nub = 1.00016 \pm  5 \, 10^{-4}$. 
    (Right)
    Temperature profiles obtained in the non-equilibrium stationary 
    states of systems in contact with stochastically thermalised
    cells at $n=0$ ($T_- = 1/2$) and $n = N+1$ ($T_+=3/2$). 
    The thick black line corresponds to the profile (\ref{noneqTn}) 
    expected from Fourier's law. Different system sizes are displayed, 
    going from $N=5$ to $N=100$. The data is barely distinguishable from
    the curve (\ref{noneqTn}) when $N$ is sufficiently large.
  }
  \label{fig.kappanoneq}
\end{figure}

A numerical computation of this quantity was performed for different 
system sizes, as shown on the left panel of figure
\ref{fig.kappanoneq}. The result of the infinite $N$ extrapolation yields
\begin{equation}
  \frac{\kappa}{\nu_\mathrm{B}} = 
  1.00016 \pm 0.0005\,.
\end{equation}

According to Fourier's law, the corresponding local temperature profile is
expected to be 
\begin{equation}
  T_n = \left[\frac{1}{2}(T_-^{\frac{3}{2}} + T_+^{\frac{3}{2}}) +
    \frac{n}{N+1} (T_+^{\frac{3}{2}} - T_-^{\frac{3}{2}})
  \right]^{\frac{2}{3}}\,,
  \label{noneqTn}
\end{equation}
which is confirmed numerically as seen in the right panel of figure
\ref{fig.kappanoneq}. 

\section{Conclusions and perspectives \label{sec.con}}

To summarize, we have successfully extended the results presented in
\cite{GG08c} to the confining dynamics of rarely interacting hard spheres.
Our results thus further validate the claim that the identity between heat
conductivity and collision frequency in such systems of confined particles
interacting through rare hard core collisions is largely independent of the
details of the underlying dynamics. 

In a forthcoming publication, we will show that the dimensionality of the
underlying dynamics is indeed arbitrary. In particular, one can consider
systems which consist of a solid structure of confining pores of arbitrary
shapes in which gas particles of arbitrary numbers are trapped. The 
geometry of the pores must be such that each isolated pore contains a fixed
number of hard spheres with micro-canonical equilibrium measure whose
energy is the sum of the kinetic energies of the gas particles. Under the
assumption that relaxation to this local equilibrium measure takes place on
time scales smaller than the time scales of energy exchanges between
neighbouring pores, a master equation of the form (\ref{mastereq}) can be
derived to describe the energy exchange process. The corresponding
stochastic kernel will in this case depend on the geometry of the pores
involved in the energy exchange and, in particular, on the precise number
of particles in those pores. The symmetry relations (\ref{symkernel}) will
thus reflect local properties of the system whose overall average yields
the corresponding macroscopic properties. In this sense, one expects that
Fourier's law can be derived in a rather general setting for such systems
of confined particles, and the value of the heat conductivity expressed in
terms of the frequency of energy exchanges between the cells.

\ack
This research is financially supported by the Belgian Federal Government
under the Inter-university Attraction Pole project NOSY P06/02 and the
Communaut\'e fran\c{c}aise de Belgique under contract ARC 04/09-312. TG is
financially supported by the Fonds de la Recherche Scientifique F.R.S.-FNRS.


\section*{References}

\begin{thebibliography}{99}

\bibitem{BLRB00}
   Bonetto F, Lebowitz J L, and Rey-Bellet L 2000 {\em Fourier Law: A Challenge
     To Theorists} in Fokas A, Grigoryan A, Kibble T, Zegarlinski B (Eds.)
   {\em Mathematical Physics 2000} (Imperial College, London).

\bibitem{GG08a}
  Gaspard G and Gilbert T  2008
  \emph{Heat conduction and Fourier's law by consecutive  local mixing and
    thermalization}
  Phys. Rev. Lett. \textbf{101} 020601.

\bibitem{GL08}
  Gilbert T and Lefevere R 2008
  \emph{Heat conductivity from molecular chaos hypothesis in locally
    confined billiard systems} 
  Phys. Rev. Lett. \textbf{101} 200601.

\bibitem{MZ83}
   Machta J and Zwanzig R 1983 
   \emph{Diffusion in a Periodic Lorentz Gas} 
   Phys. Rev. Lett. {\bf 50} 1959.

\bibitem{GG08b}
  Gaspard G and Gilbert T 2008
  \emph{Heat conduction and Fourier's law in a class of
   many particle dispersing billiards}
  New J. Phys. \textbf{10} 103004.

\bibitem{GG08c}
  Gaspard G and Gilbert T 2008
  \emph{On the derivation of Fourier's law in stochastic energy
    exchange systems} 
  J. Stat. Mech. (2008) P11021.

\bibitem{CM06}
   Chernov N and Markarian R 2006 {\em Chaotic billiards} Math.
   Surveys and Monographs {\bf 127} (AMS, Providence, RI).

\bibitem{EDHVL69}
   Ernst M H, Dorfman J R, Hoegy W R, and Van Leeuwen J M J 1969
   {\em Hard-sphere dynamics and binary-collision operators} Physica {\bf
     45} 127; Dorfman J R and Ernst M H 1989 {\em Hard-sphere
     binary-collision operators} J. Stat. Phys. {\bf 57} 581.

\bibitem{Gil76}
  Gillespie D T 1976
  \emph{General Method for Numerically Simulating the Stochastic Time
    Evolution of Coupled Chemical Reactions}
  J. Comp. Phys. \textbf{22} 403-434. 
  
\bibitem{Gil77}
  Gillespie D T 1977 
  \emph{Exact Stochastic Simulation of Coupled Chemical
    Reactions}
  J. Phys. Chem. \textbf{81} 2340-2361.


\end{thebibliography}
\end{document}